%% file: fss_v4.tex
\documentclass[showpacs,superscriptaddress,aps,twocolumn,10pt,pre]{revtex4-2}

\usepackage{amsmath,amssymb}
\usepackage{graphicx}% Include figure files
\usepackage{dcolumn}% Align table columns on decimal point
\usepackage{bm}% bold math
\usepackage{color,multirow}
\usepackage{subeqnarray}
\usepackage[colorlinks=true,%
    citecolor=blue,%
    linkcolor=blue,%
    urlcolor=blue]{hyperref}
\begin{document}
\newcommand{\blue}[1]{\textcolor{blue}{#1}}
\newcommand{\red}[1]{\textcolor{red}{#1}}
\newcommand{\green}[1]{\textcolor{green}{#1}}
\newcommand{\orange}[1]{\textcolor{orange}{#1}}

\def\today{March 10, 2025, revised: April 6, 2025}

\title{Finite-size corrections from the subleading magnetic scaling field 
for the Ising and Potts models in two dimensions}

\author{Yihao Xu}
\email{yhxu@mail.ustc.edu.cn}
\affiliation{Department of Modern Physics,
University of Science and Technology of China, Hefei,
Anhui 230026, China}
\author{Jes\'us Salas}
\email{jsalas@math.uc3m.es}
\affiliation{Universidad Carlos III de Madrid,
Departamento de Matem\'aticas,
Avenida de la Universidad 30 (edificio Sabatini),
28911 Legan\'es (Madrid), Spain}
\affiliation{Grupo de Teor\'{\i}as de Campos y F\'{\i}sica Estad\'{\i}stica,
Instituto Gregorio Mill\'an, Universidad Carlos III de Madrid,
Unidad Asociada al Instituto de Estructura de la Materia, CSIC,
Serrano 123, 28006 Madrid, Spain}
\author{Youjin Deng}
\email{yjdeng@ustc.edu.cn}
\affiliation{Department of Modern Physics, University of Science
and Technology of China, Hefei, Anhui 230026, China}
\affiliation{Hefei National Research Center for Physical Sciences
at the Microscale, University of Science and Technology
of China, Hefei 230026, China}
\affiliation{Hefei National Laboratory, University of Science and
Technology of China, Hefei 230088, China}

\date{\today}
\begin{abstract}
In finite-size scaling analyses of critical phenomena, proper consideration 
of correction terms, which can come from different sources, plays an 
important role. For the Fortuin-Kasteleyn representation of the 
$Q$-state Potts model in two dimensions, although the subleading 
magnetic scaling field, with exactly known exponent, is theoretically 
expected to give rise in finite-size-scaling analyses, 
numerical observation remains 
elusive probably due to the mixing of various corrections. We simulate 
the O($n$) loop model on the hexagonal lattice, which is in the same 
universality class as the $Q=n^2$ Potts model but has suppressed 
corrections from other sources, and provides strong numerical evidence 
for the attribution of the subleading magnetic field in finite-size 
corrections. Interestingly, it is also observed that the corrections 
in small- and large-cluster-size regions have opposite magnitudes, and, 
for the special $n=2$ case, they compensate with each other in observables 
like the second moment of the cluster-size distribution. Our finding reveals 
that the effect of the subleading magnetic field should be taken into 
account in finite-size-scaling analyses, which was unfortunately 
ignored in many previous studies.
\end{abstract}

\maketitle

%\tableofcontents

%%%%%%%%%%%%%%%%%%%%%%%%%%%%%%%%%%%%%%%%%%%%%%%%%%%%%%
% 
% INTRO 
%
%%%%%%%%%%%%%%%%%%%%%%%%%%%%%%%%%%%%%%%%%%%%%%%%%%%%%%
\section{Introduction}  \label{sec:intro}

Monte Carlo (MC) methods \cite{Madras_02,Landau_14} constitute a very 
important research tool to investigate the dynamic and static properties 
of many systems in science, and in particular, in statistical mechanics. 
In the latter field, the typical setup is a \emph{finite} system of 
dimension $d$ and linear size $L$ that undergoes a phase transition at a 
certain critical temperature. In most cases, MC simulation can only handle 
finite system, even though phase transitions only occur in the thermodynamic 
limit $N = L^d \to\infty$. 
So, in addition to the statistical error inherent to any
MC simulation, finite systems are frequently a source of systematic errors
(an exception are MC simulations of self-avoiding random walks 
or other polymer models.; see e.g., \cite{Sokal_97}.) 

Therefore, in every MC study, one has to extrapolate the results 
obtained in finite lattices to the infinite-volume limit $L\to\infty$. 
Finite-size-scaling (FSS) theory \cite{Cardy_88,Privman_90,Cardy_96}
explains how thermodynamic quantities
(like the magnetic susceptibility) behave close to a phase transition
when we take the thermodynamic limit $L\to \infty$. 

Let us consider for simplicity a finite physical system of dimension $d$, 
linear size $L$, and periodic boundary conditions. This system undergoes
a continuous phase transition that is characterized (for simplicity) by a 
single relevant thermal field $t$ (that measures the ``distance'' to the 
critical point), and an irrelevant field $u$, in the framework of 
renormalization group (RG) theory.
The singular part of the free energy density scales like 
\cite{Privman_84,Salas_00,Pelissetto_02}
\begin{equation}
f_\text{s}(g_t,g_u,L) \;=\; b^{-d}\, f_\text{s}\left( g_t\, b^{y_t}, 
g_u\,b^{y_u}, L^{-1}\, b\right) \,. 
\label{fss_fsing_RG_FSS}
\end{equation}
Here, $b$ is any positive number, $g_t$ and $g_u$ are the
nonlinear scaling fields associated to $t$ and $u$, respectively, 
and $y_t = 1/\nu >0$ and $y_u < 0$ are the 
corresponding RG eigenvalues, where $\nu$ 
is referred as the 
correlation-length exponent. Note that the inverse linear size $L^{-1}$
behaves like a relevant operator with exponent $y_L=1$. The nonlinear 
scaling fields $g_t$ and $g_u$ are analytic functions of $t$, $u$, and 
$L^{-1}$. In particular,
$g_t = a_1(u,L^{-1})\, t + a_2(u,L^{-1})\, t^2 + \cdots$, and
$g_u = b_0(u,L^{-1}) + b_1(u,L^{-1})\, t + b_2(u,L^{-1})\, t^2 + \cdots$.
Guo and Jasnow \cite{Guo_87,Guo_89} have argued, using the field-theoretic
RG, that $g_t$ and $g_u$ do not depend on $L^{-1}$. If this is the case, 
then the functions $a_i$ and $b_i$ would depend solely on $u$.  
The full free energy density is equal to
\begin{equation}
f(t,u,L) \;=\; f_\text{s}(g_t,g_u,L) + f_\text{r}(t,u) \,, 
\label{fss_f}
\end{equation} 
where the regular part $f_\text{r}$ is an analytic function of $t$ and $u$,
even at the critical point $t=0$ \cite[page 101]{Privman_91}.  

If we take the infinite-volume limit $L^{-1}=0$, and we choose $b$ such that 
$b^{y_t} \, |g_t| = 1$, we obtain 
\begin{equation}
f_\text{s}(g_t,g_u) \;=\; |g_t|^{d\nu}\, 
f_\text{s}\left( \mathrm{sign}(g_t), g_u\, |g_t|^{-y_u\nu},0\right) \,. 
\label{fss_fsing_RG}
\end{equation}
Notice that the behavior at $t=0^\pm$ might be different. 

We now require $b\, L^{-1}=1$, so that \eqref{fss_fsing_RG_FSS}
becomes
\begin{equation}
f_\text{s}(g_t,g_u,L) \;=\; L^{-d}\, 
f_\text{s}\left( g_t L^{1/\nu}, g_u L^{y_u}, 1\right) \,. 
\label{fss_fsing_RG_FSS_Bis}
\end{equation}
At criticality ($t=0$), Eqs.~\eqref{fss_f} and \eqref{fss_fsing_RG_FSS_Bis} 
reduce to 
\begin{equation}
f(0,u,L) \;=\; f_\text{r}(0,u) +  
L^{-d}\, f_\text{s}\left( 0, g_u L^{y_u}, 1\right) \,. 
\label{fss_fsing_RG_FSS_crit}
\end{equation}
The leading behavior of $f_\text{s}$ is $L^{-d}$, and we obtain 
FSS corrections of order $L^{y_u}, L^{2y_u}, \ldots$ when we expand
$f_\text{s}\left( 0, g_u L^{y_u}, 1\right)$ as a power series in $u$. 

We can obtain similar expressions for the internal energy and the specific
heat if we differentiate Eqs.~\eqref{fss_f} and \eqref{fss_fsing_RG_FSS_Bis}
with respect to $t$
once or twice, and then take the $t\to 0$ limit. In these cases, we have also
FSS corrections of order $L^{-1},L^{-2},\ldots$ due to the 
functions $a_i(u,L^{-1})$. 

The same arguments can be followed to derive the equation for the correlation
length (actually, for a definition of the correlation length that makes
sense in a finite system, e.g., the second-moment correlation length) 
\begin{equation}
\xi(t,u,L) \;=\; L\, 
\mathcal{F}_\xi \left( g_t L^{y_t}, g_u L^{y_u}\right) \,. 
\label{fss_xi_RG_FSS_Bis}
\end{equation}

Given a thermodynamic quantity $A(t)$ that diverges at the critical point
(in the infinite-volume limit) like 
\begin{equation}
A(t) \;\propto\; |t|^{-\rho}\,,  
\label{def_At}
\end{equation}
we can generalize Eq.~\eqref{fss_xi_RG_FSS_Bis} as follows 
\begin{equation}
A(t,u,L) \;=\; L^{\rho/\nu}\, 
\mathcal{F}_A\left( g_t L^{y_t}, g_u L^{y_u} \right) + A_\text{r}(t,u)\,,
\label{def_AtL}
\end{equation}
where we have added a regular background term $A_\text{r}$ 
playing a similar role as the term $f_\text{r}$
in the free energy density [cf. Eq.~\eqref{fss_fsing_RG_FSS_crit}].

Barring the effect of the irrelevant field $g_u$ in 
Eqs.~\eqref{fss_xi_RG_FSS_Bis} and \eqref{def_AtL}, 
we see that the divergence of the bulk correlation length at criticality 
is smoothed out; i.e., $\xi(0,L) \sim O(L)$, and the behavior 
of $A(t,L)$ becomes analytic in the scaled variable $g_t L^{1/\nu}$, 
even at the critical point $t=0$. FSS effects can be observed in a scaling 
window of width $O(L^{-1/\nu})$. Outside this window, $|t| \gg L^{-1/\nu}$, 
$A(t,L)$ behaves basically like $A(t)$. In this context, Li \emph{et al.}
\cite{Li_24} studied the crossover FSS theory, which shows that, as the 
critical point is approached at a slower rate with $|t| \sim L^{-\lambda}$ 
and $\lambda < 1/\nu$, the FSS becomes dependent on the parameter $\lambda$.

This description can be generalized to include more relevant and 
irrelevant fields.  
In some models, we can also find marginal operators characterized by 
zero eigenvalues $y_u=0$. These fields give rise to multiplicative 
\cite{Nauenberg_80,Cardy_80} and additive \cite{Salas_97} logarithmic 
corrections not taken into account in the previous equations. 
Logarithmic corrections can also appear when there is a ``resonance'' between
the RG eigenvalues \cite{Wegner_76}, like e.g., 
the two-dimensional (2D) Ising model. 
In this paper, we will not consider logarithmic corrections of any type. 

FSS techniques allow to compute reliable estimates of physical quantities that 
describe the infinite-volume limit (like the location of the critical point,
and its critical exponents and universal amplitudes) solely from finite-size 
data. In order to achieve this goal, one needs a good ansatz based on 
Eq.~\eqref{def_AtL}. At criticality, we have that [cf. \eqref{def_At}] 
\begin{eqnarray}
A(0,u,L) &=& L^{\rho/\nu}\, 
\mathcal{F}_A\left(0, g_u L^{y_u} \right) + A_\text{r}(0) \nonumber \\[2mm]
&\propto& L^{\rho/\nu}\, \left( 1 + a L^{y_u} + b L^{-\rho/\nu}
+  \cdots \right) \,. \qquad 
\label{def_AtL_crit}
\end{eqnarray}
It is very important to have a fairly good knowledge of the
RG exponents (i.e., $y_t$, $y_u$, etc) for the model at hand. This is 
provided in many 2D models by conformal field theory (CFT)
\cite{DiFrancesco_97}. It is also
worth noticing that FSS corrections do depend on the observable $A$: 
distinct observables may have different FSS corrections because some 
amplitudes may vanish due to symmetries. A well-known example is the 2D Ising
model. The FSS for the free energy, internal energy, and specific heat
have integer exponents \cite{Ferdinand_69,Salas_01,Salas_02}, but there is
no trace of the exponent $y_u=-4/3$ predicted by Nienhuis \cite{Nienhuis_82a}. 
In addition, some observables are 
expected to have a background term $A_\text{r}$ (e.g., the specific heat), 
while others (e.g., the correlation length) are not expected to have it. 
All these facts make the FSS analysis rather involved. 

The $Q$-state Potts model \cite{Potts_52,Wu_82,Wu_82a,Wu_84,Baxter_85} is 
an important model in statistical mechanics due to its simplicity, 
very rich phase diagram, and connections with CFT \cite{DiFrancesco_97}, 
Coulomb-gas (CG) theory \cite{Nienhuis_87}, and combinatorics 
\cite{Welsh_93}, to mention only a few. In particular, when the 2D Potts model
displays a continuous transition (i.e., when $Q\in [0,4]$), the leading and 
subleading RG exponents are known exactly in the thermal 
[cf.~\eqref{def_y_t1_and_t2}] and magnetic [cf.~\eqref{def_y_h1_and_h2}] 
sectors. This latter one appears when we add the dependence $g_h b^{y_h}$ in 
the free energy and correlation length [see 
Eqs.~\eqref{fss_fsing_RG_FSS} and \eqref{fss_xi_RG_FSS_Bis}]. 

The role of the leading ($y_{t1}$ and $y_{h1}=y_h$) and the subleading 
($y_{t2}$) RG exponents has been considered in detail by the previous 
literature. However, the effect of the subleading magnetic exponent $y_{h2}$
\eqref{def_y_h2} has been usually neglected.  
The aim of this paper is to show the effect of the subleading magnetic 
exponent $y_{h2}$ in the 2D Potts model.  

In order to achieve this goal, we have considered, instead of the standard 
spin representation of the Potts model, the so-called
Fortuin--Kasteleyn (FK) \cite{Kasteleyn_69,Fortuin_72} [cf. \eqref{def_Z_FK}].  
In this representation one can define new (geometric) observables, like the
size of the largest FK cluster, the size of the clusters, the radius of
gyration of the clusters, etc. Therefore, it is useful to introduce elements
taken from percolation theory \cite{Stauffer_94,Grimmett_99,Bollobas_06}. 
In particular, we consider the quantity $n(s;p)$---i.e., the number 
(per site) of clusters of size $s$ at probability $p$---in 
the thermodynamic limit ($L \to \infty$)
\begin{equation}
n(s;p) \;=\; s^{-\tau} \, F\left( (p-p_c)\, s^\sigma \right) \,,
\label{def_ns_ansatz_0}
\end{equation}
where $\sigma$ and $\tau$ are critical exponents and $\tau$ is usually 
refereed to be the Fisher exponent. We assume that  
$p$ is close to the critical probability $p_c$, and $s$ is large enough. 
At criticality, we obtain that $n(s;p_c) = s^{-\tau} F(0)$, 
and, taking into account corrections to scaling, 
the behavior at criticality of $n(s;p)$ should be \cite{Ziff_11,Xu_25} 
\begin{equation}
n(s;p_c) \;=\; s^{-\tau} \, \left( a + b\, s^{-\Omega} + \cdots \right) \,,
\label{def_ns_ansatz_1}
\end{equation}
where $\Omega$ is a correction-to-scaling exponent whose exact value 
is \cite{Xu_25}  
\begin{equation}
\Omega \;=\; \frac{1}{d_f \, g}\,,
\label{def_Omega_1}
\end{equation}
where 
\begin{equation}
d_f \;=\; y_{h1} \;=\; \frac{(2g+1)(2g+3)}{8g} \,,
\label{def_df}
\end{equation} 
is the fractal dimension of the FK clusters, 
and $g$ is the CG coupling that parametrizes the critical 
(or tricritical) Potts model (see below and Sec.~\ref{sec:Potts}).

In a previous paper \cite{Xu_25}, we studied the correction-to-scaling 
exponent $\Omega$ for the critical and tricritical Potts models with
$Q=1,2,3,4$. Actually, instead of simulating directly these models, we 
considered the related O$(n)$ loop model on the hexagonal lattice $G$
\cite{Nienhuis_82,Batchelor_89,Peled_19,Duminil-Copin_21}, whose 
partition function is given by 
\begin{equation}
Z_{\text{loop}}(G;x,n) \;=\; \sum_{\{\ell\}} x^{E(\ell)} \, 
n^{N(\ell)}\,, 
\label{def_Z_loop}
\end{equation}
where the sum is over all possible nonintersecting loop configurations $\ell$
on $G$, $N(\ell)$ is the number of loops in the configuration,
and $E(\ell)$ is the total length of all the loops.
The parameter $n$ represents 
the weight associated to each loop, and $x$ is a fugacity that controls
the weight of the loop's length. When $n,x$ are real
positive parameters, \eqref{def_Z_loop} has a probabilistic  
interpretation.

The free energy of the O$(n)$ loop model has been solved along two distinct
curves \cite{Nienhuis_82,Baxter_86,Baxter_87}
\begin{equation}
x_\pm \;=\; \frac{1}{\sqrt{2 \pm \sqrt{2-n}}} \,.
\label{def_xplusminus}
\end{equation}
The curve $x_+(n)$ is the critical curve of the O$(n)$ loop model and 
separates the diluted phase from the dense phase. The other curve $x_-(n)$
lies within the (critical) dense phase. This loop model has a CG 
representation \cite{Nienhuis_84,Nienhuis_87} with coupling $g$ related to
$n \in (-2,2]$ as 
\begin{equation}
n \;=\; - 2\, \cos (\pi\, g)\,, \qquad g \;\in\; \left(0,2\right]\,.
\label{def_n_vs_g}   
\end{equation}
It is known that the RG trajectories in this model keep the parameter
$n$ fixed. Moreover, the points lying on $x_+(n)$ belong to the same 
universality class as the tricritical Potts model with $Q=n^2$ states 
(and $g\in [1,2]$). In the same fashion, the points in the dense phase
[i.e., $x\in (x_+(n),\infty)$] all belong to the universality class of the
critical Potts model with $Q=n^2$ states (and $g\in (0,1]$). This happens,
in particular, to the points lying on $x_-(n)$. Notice that negative values
of $n$ can occur with the parametrization \eqref{def_n_vs_g}; but this is 
not a big issue, as $n$ is just the statistical weight of the loops in 
\eqref{def_Z_loop}. In fact, the O$(n)$ loop model with $n\in (-2,0)$ can
be studied with CFT techniques.  

Simulating the O$(n)$ loop model has several technical advantages over 
direct simulations of the critical and tricritical Potts models with $Q=n^2$:
(1) for the diluted Potts model, the \emph{exact} 
position of the tricritical point
is not known, (2) the leading thermal and magnetic exponents for the Potts 
model ($y_{t1}$ and $y_{h1}$, respectively) do not appear in the loop model
\cite{Nienhuis_87};
(3) critical slowing down (CSD) for cluster algorithms \cite{Fang_22} is
absent on $x_-$, and it is moderate on $x_+$, 
and (4) along the dense curve $x_-(n)$, the amplitudes corresponding to 
corrections from the subleading thermal scaling field vanish 
\cite{Fang_22}. In particular, there is no trace of logarithmic corrections 
for $n=2$, corresponding to the critical 4-state Potts model
\cite{Fang_22,Xu_25}. 

%%%%%%%%%%%%%%%%%%%%%%%%%%%%%%%%%%%%%%%%%%%%%%%%%%%%%%
%
% FIGURE 1
%
\begin{figure}[tbh]
\centering
\includegraphics[scale=0.9]{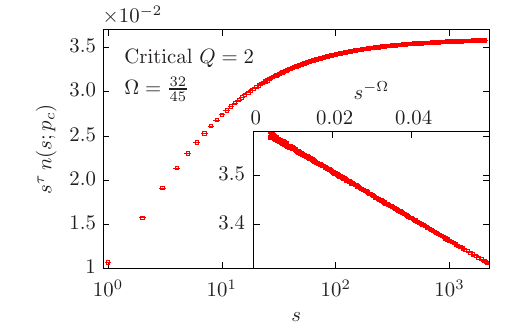}
\caption{\label{fig:ns}
Correction-to-scaling behavior of the O$(n)$ loop model on the dense 
curve $x_-$ \eqref{def_xplusminus} for $n=\sqrt{2}$; this model belongs 
to the 
universality class of the critical Ising model ($Q=2$). The numerical  
estimates were computed on a torus of linear size $L=1024$.
The main panel depicts the quantity 
$s^{\tau}\,n(s;p_c)$ vs $s$ in the range $1\le s\lesssim L$. 
The inset shows the same quantity vs $s^{-\Omega}$ (with 
$\Omega=32/45$) in the 
range $50\lesssim s \lesssim L$. A linear behavior is observed in agreement
with Eq.~\eqref{def_ns_ansatz_1}.
}
\end{figure}
%%%%%%%%%%%%%%%%%%%%%%%%%%%%%%%%%%%%%%%%%%%%%%%%%%%%%%

For each value of $n$ considered in \cite{Xu_25}, we obtained a result 
similar to that displayed in Fig.~\ref{fig:ns} for $n=\sqrt{2}$.
As $s$ increases, we see that the quantity 
$s^\tau \,n(s;p_c)$ grows and tends to a constant value $\gtrsim 3.5$. 
The transient that occurs for small and moderate values of $s$ can be 
described by a correction-to-scaling term with an exponent $\Omega$ given
by the predicted value.  

Notice that, if we express the correction term of \eqref{def_ns_ansatz_1}
in terms of a length scale $s \sim R^{d_f}$, we obtain that this correction 
term becomes $a + b' \, R^{-1/g} + \cdots$.   
So we expect that there is a FSS correction with an exponent 
$d_f \Omega = 1/g = y_{h1}-y_{h2}$. 
This argument agrees with the results of 
Refs.~\cite{Aharony_03,Asikainen_03}. They considered corrections to scaling 
of several quantities (e.g., the total mass of a FK cluster, the mass of the
hull of a cluster, etc) as a function of the radius of gyration. They found
three sources of corrections: 1) the subleading thermal exponent $y_{t2}$;
2) corrections (inspired by CG arguments) with exponents $\theta'=1/g$ (for
the total mass) or $\theta''=1/(2g)$ (for the other quantities); and 3) 
corrections with integer exponents due to ``analytic'' terms. 

Figure~\ref{fig:ns} clearly illustrates the effect of the subleading magnetic 
scaling field in the finite-cluster-size corrections for a system in the 
universality class of the 2D Ising model. 
The subleading magnetic exponent $y_{h2}$ has been computed directly by means
of MCRG methods for the 2D critical Potts model 
\cite{Rebbi_80,Swendsen_81,Swendsen_83,Shankar_85} but, to the best of our 
knowledge, a direct evidence for the existence of the corresponding FSS 
effects is still lacking. Moreover, for the 3D Ising model it has been 
argued that the subleading magnetic field is a redundant operator 
\cite{Pawley_84,Shankar_85,Baillie_92}. 
For the 2D Ising model, this issue is not clear. On one side, it has also 
been argued that this subleading magnetic field is redundant due to the 
${\mathbb Z}_2$ symmetry of the model \cite{Fang_22}.  
On the other side, two redundant operators have been found in the odd 
sector in actual MCRG simulations with $2 \times 2$ blocks
and majority rule with random tie-breakers 
\cite[and references therein]{Shankar_85}. 
However, the corresponding eigenvalues are not compatible with 
$2^{y_{h2}} = 2^{13/24} \approx 1.455653$ (see Table~\ref{tab:exponents}). 
Therefore, additional investigations will be needed to decide whether the 
subleading magnetic field is redundant or not in the 2D model.  

In the present work, we have again simulated the O$(n)$ loop model on the
hexagonal lattice using the same cluster MC algorithm as in Ref.~\cite{Xu_25}.
Our goal is to explore the contribution of the subleading magnetic 
exponent $y_{h2}$ to the finite-lattice-size corrections of that model, 
which we expect to follow a behavior of $L^{-1/g}$, where $1/g=y_{h1}-y_{h2}$. 
We have considered
two geometric quantities that belong to the ``magnetic'' sector: the mean
size of the largest cluster $C_1$ and the mean second moment $S_2$ of the 
critical cluster-size density $n(s,p_c)$ \eqref{def_ns_ansatz_1}. In fact,
we have found that both quantities display a FSS term with exponent $1/g$, 
as predicted above. In particular, for $S_2$ we argue that the $1/g$ 
contribution has two distinct origins. One of them corresponds to the 
critical cluster-size density $n(s,p_c)$ \eqref{def_ns_ansatz_1},
as illustrated in Fig.~\ref{fig:ns} for $n=\sqrt{2}$. This part accounts for
the effect of small clusters. The second origin can be attributed to the
largest cluster $C_1$, as shown in Sec.~\ref{sec:fss}. Based on the numerical
data, these two contributions share the same correction exponent $1/g$ but 
with opposite amplitudes. This leads to certain challenges in fitting the 
data for $S_2$. Furthermore, it can be observed that for $n=2$ or $Q=4$, 
although both $C_1$ and $n(s;p_c)$ exhibit such $1/g$ corrections, 
the $1/g$ correction in $S_2$ is absent, and the leading correction exponent 
is instead $2$.  

In our data analysis, we find that, in many cases, we have two FSS corrections
with similar exponents and opposite amplitudes. This is a rather unpleasant 
situation for data analysis: both terms together mimic a single correction 
term with a rather distinct exponent. In order to disentangle these two
contributions, high-precision data is needed.  

The remainder of this paper is organized as follows: 
In Sec.~\ref{sec:Potts}, we briefly review the 2D Potts
models and in Sec.~\ref{sec:fss}, we discuss the FSS corrections
in these models. Section~\ref{sec:MC} is devoted to the description of
the performed MC simulations, and in Sec.~\ref{sec:results}, 
we show our numerical findings. Finally, in Sec.~\ref{sec:discussion}, 
we present our conclusions. 

%%%%%%%%%%%%%%%%%%%%%%%%%%%%%%%%%%%%%%%%%%%%%%%%%%%%%%
% 
% Potts model 
%
%%%%%%%%%%%%%%%%%%%%%%%%%%%%%%%%%%%%%%%%%%%%%%%%%%%%%%
\section{The Potts model}  \label{sec:Potts}

In this section, we briefly describe the ``pure'' and diluted
Potts models in 2D. Let us start with the standard $Q$-state Potts model
\cite{Potts_52,Wu_82,Wu_82a,Wu_84,Baxter_85}. It can be defined on
any finite (undirected) graph $G=(V,E)$ of vertex set $V$ and edge set $E$. 
In statistical mechanics, this graph is usually chosen to be a finite 
subset of a 2D lattice with some boundary conditions (e.g., 
toroidal in MC simulations). On each vertex $x\in V$ we place
an spin $\sigma_x$ that can take $Q\in {\mathbb N}$ distinct values; that is,
$\sigma_x \in \{1,2,\ldots,Q\}$. Spins interact via a nearest-neighbor
coupling $J\in {\mathbb R}$. The Hamiltonian of this model is given by
\begin{equation}
-\beta\, \mathcal{H}_{\text{Potts}} \;=\; 
J \, \sum_{\{xy\} \in E} \delta_{\sigma_x,\sigma_y}\,,
\label{def_H_Potts}
\end{equation}
where $\delta_{a,b}$ is the Kronecker delta function, and $\beta$ is the
inverse temperature. The partition function of this model is given by
\begin{equation}
Z_{\text{Potts}}(G;Q,J) \;=\; \sum\limits_{\{\sigma\}} 
e^{-\beta\, \mathcal{H}_{\text{Potts}}}\,,
\label{def_Z_Potts}
\end{equation}
where the sum is over all possible spin configurations $\{\sigma\}$.
We will focus on the ferromagnetic regime of this model: i.e., $J>0$. 

At this stage, the parameters $Q,J$ belong to ${\mathbb N}$ and ${\mathbb R}$,
respectively. We can define the $Q$-state Potts model beyond the latter 
ranges by using the FK \cite{Kasteleyn_69,Fortuin_72} representation
\begin{equation}
Z_{\text{Potts}}(G;Q,v) \;=\; \sum\limits_{F\subseteq E} v^{|F|}\, 
Q^{k(F)}\,, 
\label{def_Z_FK}
\end{equation}
where the sum is over all spanning subgraphs $(V,F)$ of the graph
$G$, and the temperature-like variable
$v=e^J-1$ belongs to the physical interval $v\in [0,\infty)$ in the
ferromagnetic regime. The above expression is clearly a polynomial jointly
in the variables $Q,v$. Therefore, we can analytically promote these variables
from their original physical ranges to arbitrary real, or even complex, 
variables. In fact, the model \eqref{def_Z_FK} has a probabilistic 
interpretation when both variables are real and satisfy $Q,v >0$.  

The $Q$-state Potts model on any regular lattice displays a phase-transition
curve $v_c(Q)$ in the ferromagnetic regime. This phase transition is 
second-order if $Q\in [0,4]$, and first-order if $Q> 4$ \cite{Baxter_83}.  
Indeed, the form of the curve $v_c(Q)$ depends on the lattice structure,
but the critical behavior is universal. In particular, the critical 
$Q$-state Potts model can be represented as a CG with 
parameter $g$ \cite{Nienhuis_84,Nienhuis_87}. The relation between $Q$ and 
$g$ is given by 
\begin{equation}
\sqrt{Q} \;=\; - 2\, \cos (\pi\, g)\,, \quad g \in \left(0,1\right]\,. 
\label{def_Q_vs_g}   
\end{equation}
Note that we find $\sqrt{Q}<0$ for $g\in(0,1/2)$. Although this may look 
unphysical, it has no practical importance as the partition function 
\eqref{def_Z_FK} depends on $Q = 4\cos^2(\pi g) \ge 0$. 

\def\kk{\phantom{-}}
\def\kz{\phantom{0}}
%%%%%%%%%%%%%%%%%%%%%%%%%%%%%%%%%%%%%%%%%%%%%%%%%%%%%%
%
% TABLE 1
%
\begin{table*}[bt]
\caption{\label{tab:exponents}
Critical exponents for the critical and tricritical $Q$-state Potts models.
For each model and value of $Q$, we show the CG coupling $g$, the dominant
thermal exponents $y_{t1}$ \eqref{def_y_t1}, 
the subdominant thermal exponent $y_{t2}$ \eqref{def_y_t2},
the dominant magnetic exponent $y_{h1}$ \eqref{def_y_h1} [which is 
equal to the fractal dimension $d_f$], 
the subdominant magnetic exponent $y_{h2}$ \eqref{def_y_h2}, and  
their difference $y_{h1}-y_{h2}=1/g$. 
}
\begin{ruledtabular}
\begin{tabular}{lcccccccc}
	&    &\multicolumn{3}{c}{Critical} &   
	&\multicolumn{3}{c}{Tricritical} \\
	\cline{3-5} \cline{7-9} \\[-3mm] 
	&    &$Q=1$  &$Q=2$  &$Q=3$  &$Q=4$  &$Q=3$      &$Q=2$      &$Q=1$\\
	&$g$ &$2/3$  &$3/4$  &$5/6$  &1      &$7/6$      &$5/4$      &$4/3$ \\
	$y_{t1}$&$3-3/2g$
	&$3/4$  &1      &$6/5$  &$3/2$  &$12/7$     &$9/5$      &$15/8$\\
	$y_{t2}$&$4-4/g$ 
	&$-2$\kk&$-4/3\kk$ &$-4/5\kk$ & 0     &$4/7$      &$4/5$      &1\\
	$y_{h1}=d_f$ &$(2g+1)(2g+3)/8g$ 
	&$91/48$ &$15/8$ &$28/15$&$15/8$&$40/21$    &$77/40$    &$187/96$\\
	$y_{h2}$ &$(2g-1)(2g+5)/8g$ 
	&$19/48$ &$13/24$&$2/3$  &$7/8$  &$22/21$    &$9/8$      &$115/96$
	\\\\[-2mm]
	$y_{h1}-y_{h2}$ &$1/g$  
	&$3/2$   &$4/3$  &$6/5$  &$1$    &$6/7$      &$4/5$      &$3/4$
\end{tabular}
\end{ruledtabular}
\end{table*}
%%%%%%%%%%%%%%%%%%%%%%%%%%%%%%%%%%%%%%%%%%%%%%%%%%%%%%

A natural generalization of the above model is the diluted $Q$-state Potts
model. The basic idea is to allow for vacancies in the graph $G$, which can
be represented by integer variables $\tau_x \in \{0,1\}$ placed on the
vertices of the lattice. In particular, $\tau_x=0$ ($\tau_x=1$)
means that the corresponding vertex $x\in V$ is empty (occupied). 
The Potts Hamiltonian \eqref{def_H_Potts} can be generalized in several ways. 
One simple form is \cite{Nienhuis_79,Nienhuis_80,Wu_82} 
\begin{equation}
-\beta\, \mathcal{H}_{\text{dP}} \;=\; 
\sum_{\{x,y\}\in E} \tau_x \tau_y (K + J \delta_{\sigma_x,\sigma_y}) 
- \Delta \sum_{x\in V} \tau_x \,. 
\label{def_H_diluted_Potts}
\end{equation}
In this equation, $\Delta$ plays the role of the chemical potential 
governing the concentration of vacancies. Other Hamiltonians have been
proposed in the literature: see e.g. 
\cite{Murata_79,Nienhuis_82a,Janke_04,Deng_05,Qian_05}. In some cases,
it is possible to find a FK-type representation.  

The diluted Potts model appears naturally when we perform a 
RG transformation on the pure Potts model 
\eqref{def_H_Potts} \cite{Nienhuis_79,Nienhuis_80}.
In particular, the variable $Q$ remains constant under RG. 
On the critical surface for $Q\in [0,4]$,
there is a line of attractive critical fixed points (in the
same universality class as the pure Potts model), and there is another
line of (repulsive) tricritical fixed points (belonging to a new  
universality class). Both lines meet at $Q=4$. For $Q> 4$, 
the system renormalizes to a discontinuity fixed point (at zero temperature),
as expected \cite{Nienhuis_75,Klein_76,Fisher_78}.

There is also a CG representation of the tricritical Potts model for
$Q\in [0,4]$. In particular, the relation between $Q$ and $g$ is also
given by \eqref{def_Q_vs_g}, but the range for $g$ is now $g\in[1,2]$. 
In this sense, the tricritical Potts model is the analytic extension of 
the critical one, or \emph{vice versa}.  

The leading and subleading thermal exponents $y_{t1}$ and
$y_{t2}$ relate to $g$ as \cite{Nienhuis_87}
\begin{subeqnarray}
\slabel{def_y_t1}
y_{t1} &=& \frac{3\, (2g-1)}{2\, g} \,, \\
y_{t2} &=& \frac{4\, (g-1)}{g} \,,
\slabel{def_y_t2}
\label{def_y_t1_and_t2}
\end{subeqnarray}
and the corresponding magnetic exponents $y_{h1}$ and $y_{h2}$ are given by
\begin{subeqnarray}
\slabel{def_y_h1}
y_{h1} &=& \frac{(2g+1)\, (2g+3)}{8\, g}\,, \\
y_{h2} &=& \frac{(2g-1)\, (2g+5)}{8\, g} \,.
\slabel{def_y_h2}
\label{def_y_h1_and_h2}
\end{subeqnarray}
The exponents \eqref{def_y_t1_and_t2} and \eqref{def_y_h1_and_h2}
are displayed in Table~\ref{tab:exponents} for future use.
The subleading thermal exponent \eqref{def_y_t2} corresponds to the 
dilution operator. This one is relevant for the tricritical Potts model and
irrelevant for the critical Potts model.
In the critical 4-state Potts model, corresponding to $g=1$, that operator
is marginal with $y_{t2}=0$. More precisely, the dilution operator
is marginally irrelevant at $Q=4$. This is the origin of multiplicative
\cite{Nauenberg_80,Cardy_80} and additive \cite{Salas_97} logarithmic
corrections.

From the leading eigenvalues \eqref{def_y_t1} and \eqref{def_y_h1}, 
one can obtain the standard critical exponents: e.g., 
\begin{subeqnarray}
\nu &=& \frac{1}{y_{t1}} \;=\; \frac{2\,g}{3\, (2g-1)} \,,  \\[2mm]
\gamma &=& \frac{2y_{h1}-d}{y_{t1}} \;\;\stackrel{d=2}{=}\;\; 
\frac{3+4g^2}{6\, (2g-1)}\,,
\end{subeqnarray}
and the rest can be derived using the hyperscaling relations. 

%%%%%%%%%%%%%%%%%%%%%%%%%%%%%%%%%%%%%%%%%%%%%%%%%%%%%%
% 
% FSS 
%
%%%%%%%%%%%%%%%%%%%%%%%%%%%%%%%%%%%%%%%%%%%%%%%%%%%%%%
\section{Finite-size scaling}  \label{sec:fss}

In percolation theory, the fraction of the largest cluster over the 
system volume $C_1/L^d$ 
acts as an order parameter, and the second moment $S_2$ of the cluster 
size distribution $n(s;p_c)$ corresponds to the magnetic susceptibility. 
According to the standard FSS theory, 
the behavior of $C_1$ [cf. \eqref{def_C1}]
at criticality can be obtained by differentiating 
the free energy \eqref{fss_fsing_RG_FSS_Bis} with respect to the magnetic 
scaling field, leading to 
\begin{eqnarray}
C_1 &=& c_0+a_1 L^{y_{h1}} + a_2  L^{y_{h2}} + \cdots  \nonumber \\[2mm] 
&\propto&  L^{y_{h1}} \, \left( 1 + a'  L^{y_{h2} - y_{h1}} + 
           c_0' L^{-y_{h1}} \cdots \right)\,, \qquad  
\label{def_fss_C1_1} 
\end{eqnarray}
where $C_1$ is a shorthand for $C_1(p_c;L)$, and the background term 
$c_0$ comes from the analytical part of the free energy.  
Similarly, the FSS behavior of $S_2$ [cf. \eqref{def_obs_S2}/\eqref{def_S2}]
at criticality can be obtained by 
differentiating twice the free energy with respect to the magnetic 
scaling field, which gives 
\begin{eqnarray}
S_2 &=& s_0+b_1 L^{2y_{h1}-d} + b_2  L^{y_{h1}+y_{h2}-d} + \cdots  
\nonumber \\[2mm]
&\propto&  L^{\gamma/\nu} \, \left( 1 + b' L^{-1/g} + 
     s_0' L^{-\gamma/\nu} + \cdots \right)\,, 
\label{def_fss_S2_1} 
\end{eqnarray}
where $S_2$ is a shorthand for $S_2(p_c;L)$, and $\gamma/\nu=2y_{h1}-d$. 
Hence, the first correction term should be of order $y_{h2}-y_{h1}=-1/g$ 
in agreement with Refs.~\cite{Aharony_03,Asikainen_03}.

The basic FSS behavior of $S_2$ \eqref{def_fss_S2_1} can be deduced 
alternatively by using
Eqs.~\eqref{def_ns_ansatz_1}, \eqref{def_Omega_1}, and~\eqref{def_fss_C1_1} 
\begin{eqnarray}
S_2
&=&\int_{1}^{C_1} s^2\, n(s;p_c)\, \mathrm{d} s 
\nonumber \\[2mm] 
&\propto& L^{\gamma/\nu}\, \left(1+ b' L^{-1/g} + c' L^{-\gamma/\nu}+\cdots 
\right)\,,
\label{eq_fss_S2}	
\end{eqnarray}
where we have made use of the relations $\tau = 1+d/d_f$ and 
$\gamma/\nu=2y_{h1}-d$. 
In other words, the \emph{leading} FSS behavior of $S_2$ can be 
obtained by differentiating the free energy or integrating the  
cluster-number density. Both methods yield, as expected, the same form of 
the leading FSS corrections $\sim L^{-1/g}$.
Furthermore, by using the second procedure, it is clear that the 
correction amplitude $b'$ in $S_2$ \eqref{eq_fss_S2} has contributions 
from both the largest cluster $C_1$ and the smaller ones [via $n(s;p_c)$].

Let us now discuss the FSS corrections that may arise from other 
sources:
1) The background terms in Eqs.~\eqref{def_fss_C1_1} and \eqref{def_fss_S2_1}
($c_0$ and $s_0$, respectively) are expected to exist and they act 
effectively as correction terms with exponents $-y_{h1}$ for $C_1$ and 
$d-2y_{h1}=-\gamma/\nu$ for $S_2$.
2) One should have a term $\propto L^{2y_{h2}-d}$ for $S_2$, not explicitly 
shown in Eq.~\eqref{def_fss_S2_1},
which would give a correction term with exponent $-2/g$.
3) For the critical Potts model, a typical source of FSS corrections 
is the subleading thermal scaling field, which is irrelevant 
and has exponent an $y_{t2} < 0$ (see Table~\ref{tab:exponents}).
Fortunately, exactly along the dense branch $x_-(n)$ of the O($n$) loop model, 
the amplitude of the subleading thermal scaling field is exactly zero, 
and thus the correction term with exponent $y_{t2}$ is expected to be absent, 
as numerically confirmed in previous studies \cite{Fang_22}. 

\begin{figure}[tbh]
\centering
\includegraphics[scale=0.95]{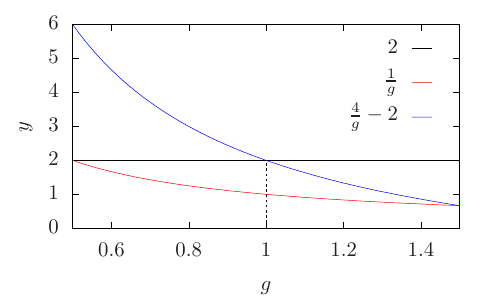}
\caption{\label{fig:exponents}
Correction-to-scaling exponents in Eq.~\eqref{def_expansion_Pi}. We
show the behavior of the different exponents appearing in 
Eq.~\eqref{def_expansion_Pi} as a function of $g\in [1/2,3/2]$ 
(i.e., for $Q\in [0,4]$). 
The exponent $1/g$ (red curve) is the most relevant in the
interval $g\in [1/2,3/2]$. The second most relevant exponent is 
$2$ (black horizontal line) in the interval $g\in[1/2,1]$, and $4/g-2$
(blue curve), in the interval $g\in[1,3/2]$. 
}
\end{figure}

Additional insights can be obtained by following the argument shown in 
Sec.~III of Ref.~\cite{Xu_25}.
Consider an annulus of size $R_1 \times R$ with periodic (free)
boundary conditions on the second (first) coordinate.
When the inner radius $R_1$ is kept finite while the outer radius 
becomes asymptotically large ($R \gg 1$),
the crossing probability $\Pi(R)$ that a cluster connects the two boundaries is
given by  
\begin{multline}
\Pi(R) \;=\; R^{d_f -2} \,
\left( A_0 + A_1 \, R^{-1/g} 
+ A_2 \, x^{-2} \right. \\[2mm]  
\left.      + A_3 \, R^{-(4/g-2)}  
+ A_4 \, R^{-(1/g +2)} + \ldots  \right) \,,
\label{def_expansion_Pi}
\end{multline}
where the dots stand for higher-order corrections and
the leading correction exponent $-1/g$ agrees with the previous analyses.
Figure~\ref{fig:exponents} shows the exponents in Eq.~\eqref{def_expansion_Pi}
as a function of $g \in [1/2,3/2]$, which corresponds to $Q\in[0,4]$. 

%%%%%%%%%%%%%%%%%%%%%%%%%%%%%%%%%%%%%%%%%%%%%%%%%%%%%%
%
% FIGURE 2
%
%%%%%%%%%%%%%%%%%%%%%%%%%%%%%%%%%%%%%%%%%%%%%%%%%%%%%%

As shown in Ref.~\cite{Xu_25}, the quantity $\Pi(R)$ \eqref{def_expansion_Pi}
can be related to the probability at criticality $P_{\ge R}$ that an 
occupied vertex is connected to a FK cluster of size greater than or equal to
$s\sim R^{d_f}$. In fact, $P_{\ge s}$ is given by the first moment of the 
cluster density at criticality $n(s;p_c)$, so we expect that the FSS 
terms that appear in Eq.~\eqref{def_expansion_Pi} will also appear in 
the FSS behavior of both $C_1$ and $S_2$.  

To summarize, let us write the FSS behavior of both $C_1$ and $S_2$ as
\begin{equation}
A \;=\; L^{y_A}\, \left( a + b L^{-y_1} + c L^{-y_2} + \cdots \right) 
\end{equation}
where $y_{C_1}=y_{h1}=d_f$, $y_{S_2}=2y_{h1}-d$, and 
$0 < y_1 < y_2$ are the first two dominant exponents. 
Based in the above discussion, for both quantities the leading 
FSS correction exponent is 
\begin{equation}
y_1 \;=\;  \frac{1}{g} \,. 
\end{equation}
The exact values of this exponent for the critical and tricritical Potts 
models are listed in Table~\ref{tab:exponents}. 
For the subdominant correction, the exponent is given by 
\begin{equation}
y_2 = \min\left(2, \frac{4}{g}-2 , y_{h1}\right) \, 
\end{equation}
for $C_1$, and 
\begin{equation}
y_2 = \min\left(2, \frac{4}{g}-2 , \frac{2}{g}, 2-2y_{h1}\right) \, 
\end{equation}
for $S_2$ (where $d=2$ has been used). 

%%%%%%%%%%%%%%%%%%%%%%%%%%%%%%%%%%%%%%%%%%%%%%%%%%%%%%
% 
% MONTE CARLO 
%
%%%%%%%%%%%%%%%%%%%%%%%%%%%%%%%%%%%%%%%%%%%%%%%%%%%%%%
\section{Monte Carlo simulations}  \label{sec:MC}

In this section we describe the MC simulations we have performed
in this work. We have used a cluster MC algorithm to simulate the O$(n)$ loop 
model on both branches $x_\pm$ \eqref{def_xplusminus}. 
Instead of simulating this model, we have 
used its representation as a generalized Ising model on the dual 
triangular lattice. We refrain from giving details of the 
algorithm, as it is well explained in Refs.~\cite{Xu_25,Fang_22}. 

As explained in the previous section, we have focused on two main physical 
quantities. In each iteration of the MC 
simulation, we recorded the sizes of the different FK clusters in the 
system. Let us denote $\mathcal{C}_k$ the size of the $k$th largest cluster. 

The first quantity of interest is the mean size of the largest cluster $C_1$
\begin{equation}
C_1 \;=\; \langle \mathcal{C}_1 \rangle \,.
\label{def_C1} 
\end{equation} 

The second quantity measured is the mean second moment $S_2$ of the 
critical cluster size $n(s;p_c)$. We first define the observable 
$\mathcal{S}_2$
\begin{equation}
\mathcal{S}_2 \;=\; \frac{1}{L^d} \sum_i \mathcal{C}_i^2 \,,
\label{def_obs_S2}
\end{equation}
where the sum is over all clusters in the system. Then, $S_2$ is given by 
\begin{equation}
S_2 \;=\; \langle \mathcal{S}_2 \rangle\,.
\label{def_S2}
\end{equation}

We have simulated the O$(n)$ loop model at $n=1,\sqrt{2},\sqrt{3},2$ on the
dense branch $x_-$. These models have the same universality class as the
corresponding critical Potts model with $Q=n^2$ states. 
We have also simulated the O$(n)$ loop model at $n=1,\sqrt{2},\sqrt{3}$ 
on the dilute branch $x_+$. In this case, these models have the same
universality class than the corresponding tricritical Potts model with 
$Q=n^2$ states. 
For each model, the systems had linear sizes equal to  
$L=4$, 5, 6, 7, 8, 9, 10, 11, 12, 14, 16, 18, 20, 24, 28, 32, 36, 40, 48, 56, 
64, 80, 96, 112, 128, and 256. 
For the O$(n)$ loop model at $n=\sqrt{2},\sqrt{3}$ on the dense branch 
$x_-$, we have also considered systems of linear size $L=1024$.
We have used periodic boundary conditions in all our simulations.  
Note that, since we are mainly interested in corrections to scaling rather
than in the leading scaling term, simulations for small system sizes 
play a significant role, and it is more important to achieve 
high-precision data for small and moderate values of $L$ than going to even 
larger system sizes.

For systems with $L \leq 256$, more than $5 \times 10^7$ 
statistically independent samples were generated, while for 
the system with $L=1024$, more than $5 \times 10^6$ independent 
samples were generated.   

%%%%%%%%%%%%%%%%%%%%%%%%%%%%%%%%%%%%%%%%%%%%%%%%%%%%%%
%
% RESULTS
%
%%%%%%%%%%%%%%%%%%%%%%%%%%%%%%%%%%%%%%%%%%%%%%%%%%%%%%
\section{Results} \label{sec:results}

In this section, we discuss our findings. We have measured the quantities
$C_1(L)$ and $S_2(L)$ on many finite lattices of linear size $L$ and periodic
boundary conditions. 
In order to study the thermodynamic limit ($L\to\infty$), we need to perform 
least-squares fits to the nonlinear FSS ansatz
\begin{equation}
A(L) \;=\; L^{y_A} \, \Big(a + b_1\, L^{-\omega} + b_2\, L^{-y_2} + 
b_3\, L^{-y_3} \Big)
\label{def_ansatz}
\end{equation}
where $0<\omega < y_2 < y_3$ are the correction-to-scaling exponents. 

We have seen in Sec.~\ref{sec:fss}, that for $A=C_1$, $y_{C_1}=d_f$, and
for $A=S_2$, $y_{S_2}=2d_f-d=2d_f-2$. As the fractal dimension $d_f$ is 
known for the Potts model [cf. Eq.~\eqref{def_df}], we consider
the reduced quantities
\begin{subeqnarray} 
\slabel{def_C1_tilde}
\widetilde{C}_1 &=& C_1\, L^{-d_f} \,,  \\[2mm]
\widetilde{S}_2 &=& S_2\, L^{-(2d_f-2)} \,. 
\slabel{def_S2_tilde}
\label{def_obs_tilde}
\end{subeqnarray} 
These quantities behave as follows: 
\begin{equation}
\widetilde{A}(L) \;=\; a + b_1\, L^{-\omega} + b_2\, L^{-y_2} + 
b_3\, L^{-y_3}\,. 
\label{def_ansatz_OK}
\end{equation}
The parameter $a$ gives the value of $\widetilde{A}$ in the
thermodynamic limit $L\to\infty$. The parameters $b_i$ represent the amplitudes
of the corresponding FSS correction terms.
The notation of this equation will be used when we discuss the numerical 
results of the MC simulations. 

We have used {\sc Mathematica}'s built-in function {\tt NonlinearModelFit} to 
perform the weighted least-squares method for both linear and nonlinear
fits to the ansatz \eqref{def_ansatz_OK}. 
In order to detect corrections to scaling not taken 
into account in the ansatz \eqref{def_ansatz_OK}, we have repeated 
each fit by only allowing data with $L\ge L_\text{min}$. We then study 
the behavior of the estimated parameters as a function of $L_\text{min}$. 
In general, our preferred fit will correspond to the smallest $L_\text{min}$
for which the goodness of fit is reasonable,  
and for which subsequent increases in $L_\text{min}$
do not cause the $\chi^2$ to drop \emph{vastly} more than one unit per 
degree of freedom.  
For each fit we report the observed value of the $\chi^2$, and the
number of degrees of freedom (DF).

In the next sections we will discuss the simulated models ordered in
increasing value of the CG coupling $g$.  

%%%%%%%%%%%%%%%%%%%%%%%%%%%%%%%%%%%%%%%%%%%%%%%%%%%%%%
%
% Q=1,2,3
%
%%%%%%%%%%%%%%%%%%%%%%%%%%%%%%%%%%%%%%%%%%%%%%%%%%%%%%

%%%%%%%%%%%%%%%%%%%%%%%%%%%%%%%%%%%%%%%%%%%%%%%%%%%%%%
%
% FIGURE 3
%
\begin{figure}[t]
\includegraphics[scale=0.9]{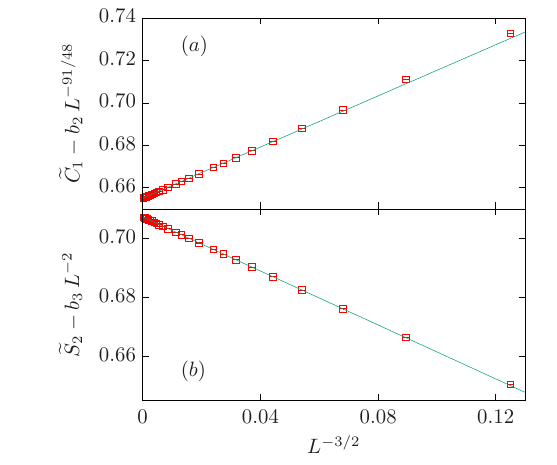}
\caption{\label{fig:Q1} 
FSS corrections for the O(1) loop model on the dense branch, which belongs
to the same universality class as the critical one-state Potts model. 
We show the quantities $\widetilde{C}_1 - b_2 L^{-91/48}$ with $b_2=-0.582$ 
in panel (a), and $\widetilde{S}_2 - b_3 L^{-2}$ with $b_3=0.679$ in panel (b).
In both panels, data is plotted vs $L^{-3/2}$. 
The smallest three system sizes are $L=4,5,6$, and the corresponding 
data points fall approximately on the straight line from the fits. This 
is a rather surprising fact.
}
\end{figure}
%%%%%%%%%%%%%%%%%%%%%%%%%%%%%%%%%%%%%%%%%%%%%%%%%%%%%%

\subsection{\texorpdfstring{%
	O$\bm{(n)}$ loop model on the dense branch $\bm{x_-}$} 
              {O(n) loop model on the dense branch}} \label{sec:Q=1-3}

%%%%%%%%%%%%%%%%%%%%%%%%%%%%%%%%%%%%%%%%%%%%%%%%%%%%%%
% 
% TABLE 2 for Q=1 
%
\begin{table*}[tb] 
\caption{\label{tab:Q=1}
FSS corrections for the O$(1)$ loop model on the dense branch $x_-$, 
which belongs to the same
universality class as the critical one-state Potts model. Both models are
characterized by $g=2/3$, $d_f=91/48$, and $2d_f-2 = 43/24$. 
We show the fits of the quantities $\widetilde{C}_1$ and $\widetilde{S}_2$ 
to the full ansatz \eqref{def_ansatz_OK}.
}
\begin{ruledtabular}
\begin{tabular}{ccllllllll}
&\multicolumn{9}{c}{Fits for $n=1$ on the dense branch $x_-$} \\
\multicolumn{1}{c}{$\widetilde{O}$} 	
&\multicolumn{1}{c}{$L_{\rm min}$}
&\multicolumn{1}{c}{$a$} 
&\multicolumn{1}{c}{$b_1$}
&\multicolumn{1}{c}{$\omega$}
&\multicolumn{1}{c}{$b_2$}
&\multicolumn{1}{c}{$y_2$}  
&\multicolumn{1}{c}{$b_3$}  
&\multicolumn{1}{c}{$y_3$}  
&\multicolumn{1}{c}{$\chi^2/{\rm DF}$} \\
\hline\\[-3mm]
$\widetilde{C}_1$ 
&28 &0.655033(7) &\kk0.304(9) &1.383(9)&\kk0     &      &\kk0     &  &6.9/9\\ 
&12 &0.655045(6) &\kk0.60(4)	   &1.50(1)	&-0.57(6) &91/48 &\kk0 &  &10.0/14\\ 
&16 &0.655052(8) &\kk0.58(7)  &1.50(2) &\kk0     &      &-0.7(1)  &2 &7.1/12\\ 
&12 &0.655045(6) &\kk0.61(2)  &  3/2   &-0.6(3)  &91/48 &\kk0.1(3)&2 &10.0/14\\ 
&12 &0.655046(3) &\kk0.602(2) &  3/2   &-0.582(6)&91/48 &\kk0     &  &10.0/15\\\\[-2mm]
$\widetilde{S}_2$
&24   &0.70729(1)  &-0.175(10)&1.31(2) &\kk0       &      &\kk0       &  &7.5/10\\ 
&\kz7 &0.707245(9) &-0.67(4)  &1.55(1) &\kk0.87(5) &43/24 &\kk0       &  &12.0/19\\ 
&\kz7 &0.707251(9) &-0.46(2)  &1.50(1) &\kk0       &      &\kk0.68(3) &2 &11.4/19\\ 
&\kz7 &0.707250(8) &-0.46(2)  &3/2     &\kk0.00(8) &43/24 &\kk0.68(7) &2 &11.4/19\\
&\kz7 &0.707250(5) &-0.458(1) &3/2     &\kk0       &      &\kk0.679(3)&2 &11.4/20\\
\end{tabular}
\end{ruledtabular}
\end{table*}
%%%%%%%%%%%%%%%%%%%%%%%%%%%%%%%%%%%%%%%%%%%%%%%%%%%%%%

In this subsection we will discuss our results for the O$(n)$ loop model on
the dense branch $x_-$ with $n=1,\sqrt{2},\sqrt{3}$. These models belong to
the same universality class as the critical Potts models
with $Q=1,2,3$ states, respectively. 
The results are rather similar qualitatively, so we will discuss the first case 
in more detail, and be brief in the other two cases.  

%%%%%%%%%%%%%%%%%%%%%%%%%%%%%%%%%%%%%%%%%%%%%%%%%%%%%%
%
% Q=1
%
%%%%%%%%%%%%%%%%%%%%%%%%%%%%%%%%%%%%%%%%%%%%%%%%%%%%%%
\subsubsection{\texorpdfstring{$n=1$}{n=1}} \label{sec:Q=1}

Let us start with the O$(1)$ loop model on the dense branch, 
which reduces to the triangular-lattice site percolation and 
belongs to the same universality class as the critical 
one-state Potts model---i.e., the percolation universality. 
It is characterized by $g=2/3$, 
$d_f=91/48 \approx 1.895833$, and $2d_f-2 = 43/24 \approx 1.791667$.

In the least-squared fit, we include at most three correction terms as 
in Eq.~\eqref{def_ansatz_OK}, and, furthermore, fix $y_3=2$ if not being 
explicitly specified. This is because, despite of our extensive simulations, 
the precision of our data is not sufficient to simultaneously discern several 
correction terms. In addition, it seems unnecessary to include more rapidly 
decaying corrections.  

We start with the reduced largest-cluster size $\tilde{C}_1$.  The first 
step consists in fitting the data to a single power-law 
[i.e. we set $b_2=b_3=0$ in \eqref{def_ansatz_OK} so there are three free 
parameters $\{a,b_1,\omega\}$]. We obtain a sensible fit for $L_\text{min}=28$, 
giving $\omega=1.383(8)$ with $\chi^2/\text{DF}=6.9/9$
(see first data row of Table~\ref{tab:Q=1}). 
Even though the fit looks reasonable, the estimate 
for $\omega$ is significantly away the expected result $1/g=3/2$. 
We will show below that this is due to
the effect of neglecting higher-order FSS corrections in the ansatz.

In a second step, we try to perform a fit to the ansatz  
\eqref{def_ansatz_OK} with $b_3=0$; i.e., with five free parameters 
$\{a,b_1,b_2,\omega,y_2\}$. In this case, we do not obtain any good fit:
either the program is unable to find a solution and/or the estimates for
the parameters and their error bars take unreasonably large values. 
A similar behavior is found if we fix $\omega=1/g=3/2$ and $b_3=0$ in the 
ansatz \eqref{def_ansatz_OK} so that there are four free parameters
$\{a,b_1,b_2,y_2\}$. Again, the program is unable to converge 
or the estimates are unreasonably large. 
In other words, our data are not sufficiently accurate to simultaneously 
distinguish and determine two correction exponents whose values are 
rather close to each other.
We refrain to include these failed fits in Table~\ref{tab:Q=1}. 

The third step involves performing a fit with two correction terms, where the 
subleading correction exponent is fixed. For $y_2 = d_f = 91/48$ and $b_3 = 0$,
a good fit is achieved for $L_\text{min} = 12$ by using a four-parameter fit 
with $\{a, \omega, b_1, b_2\}$, yielding $\omega = 1.50(1)$. Similarly, 
for $y_3 = 2$ and $b_2 = 0$, a good fit is obtained for $L_\text{min} = 16$ 
with a four-parameter fit involving $\{a, \omega, b_1, b_3\}$, resulting 
in $\omega = 1.50(2)$ (see second and third data row on Table~\ref{tab:Q=1}). 
Both fits show excellent agreement with the theoretical expectation 
$1/g = 3/2$. It is noted that the two amplitudes are similar in magnitude 
but have opposite signs in both fits.

To get further indication about which subleading correction term is somewhat 
more probable, we perform a fit with three correction terms for which all 
the correction exponents are fixed as $\omega=1/g=3/2$, $y_2=91/48$ and 
$y_3=2$. Our preferred fit is obtained for $L_\text{min}=12$. 
The amplitude $b_3$ is zero within error bars: i.e., $b_3=0.1(3)$. 
Therefore, we conclude that we need only two exponents $\omega=3/2$ and 
$y_2=d_f=91/48$ to give account of the data with $L\ge L_\text{min}=12$. 
The last row of the $\widetilde{C}_1$ block in Table~\ref{tab:Q=1} displays 
the results of this final fit. These results
are used in Fig.~\ref{fig:Q1}(a), which depicts the quantity 
$\widetilde{C}_1-b_2L^{-91/48}$ vs $L^{-3/2}$ where $b_2=-0.582$ is taken 
from the fit. According to our discussion, the data points form a straight 
line with a positive slope. 

We now see why the first fits are not successful: there are two FSS 
corrections with exponents not too different (namely, $1/g=3/2=1.5$ and 
$y_2=d_f=91/48\approx 1.895833$), and almost opposite amplitudes 
$b_1=0.602(2)$ vs $b_2=-0.582(6)$. This is really a hard scenario for
estimating these parameters, and some theoretical input is needed to
disentangle these two contributions. Actually, they merge in such a way
that they mimic a single power-law, like the one we obtained in the first 
step. 

We will see that this subtle scenario persists for the second moment of the 
cluster sizes $S_2$ and for the other values of $n$.

In summary, the MC data for $\widetilde{C}_1$ can be described with two FSS 
corrections: $\omega=1/g$ and $y_2=d_f$. The amplitude of the former 
(latter) correction term is positive (negative).  

We can follow the same procedure for the $\widetilde{S}_2$ data (see
Table~\ref{tab:Q=1}). If we set $b_2=b_3=0$, we obtain a good fit for 
$L_\text{min}=24$ yielding $\omega=1.31(2)$. Again, this latter 
estimate is very far from the expected one $1/g=3/2$.  

If we fix $b_2$ or $b_3$ to 0 with $y_2 = 2d_f - 2 = 43/24$ and $y_3=2$, 
we obtain sensible fits for
$L_\text{min}=7$. The corresponding estimates for $\omega$ are similar: 
$1.55(1)$ ($1.50(1)$) for $b_3=0$ ($b_2=0$). The corresponding 
amplitudes are positive, while $b_1$ is negative (see Table~\ref{tab:Q=1}).  

If $\omega = 3/2$, $y_2 = 2d_f-2$, and $y_3 = 2$ are fixed, with $b_2$ and 
$b_3$ treated as free parameters, we obtain a good fit for $L_\text{min} = 7$. 
It is worth noticing that the amplitude $b_2=0.00(8)$ is zero within errors, 
so we can safely assume $b_2=0$. 
Finally, if we fit the data to the ansatz \eqref{def_ansatz_OK} with 
$\omega=3/2$, $y_3=2$, and $b_2=0$, we get a nice result for 
$L_\text{min}=7$. In this case, the amplitudes $b_1$ and $b_3$
are again similar with opposite signs. 
Figure~\ref{fig:Q1}(b) shows the quantity $\widetilde{S}_2 - b_3 L^{-2}$ vs 
$L^{-3/2}$, and the data points form a straight line with negative slope.  

Therefore, the MC data can be described again with two FSS corrections:
$\omega=1/g$, and $y_3=2$. The amplitude of the former (latter)
correction term is negative (positive). 

%%%%%%%%%%%%%%%%%%%%%%%%%%%%%%%%%%%%%%%%%%%%%%%%%%%%%%
%
% Q=2
%
%%%%%%%%%%%%%%%%%%%%%%%%%%%%%%%%%%%%%%%%%%%%%%%%%%%%%%
\subsubsection{\texorpdfstring{$n=\sqrt{2}$}
{n=sqrt{2}}} \label{sec:Q=2}

The O($\sqrt{2})$ loop model on the dense branch belongs to the same 
universality class as the critical Ising model. They are characterized by 
$g=3/4$, $d_f=15/8 = 1.875$, and $2-2d_f=7/4=1.75$.   

%%%%%%%%%%%%%%%%%%%%%%%%%%%%%%%%%%%%%%%%%%%%%%%%%%%%%%
%
% FIGURE 4
%
\begin{figure}[t]
\includegraphics[scale=0.9]{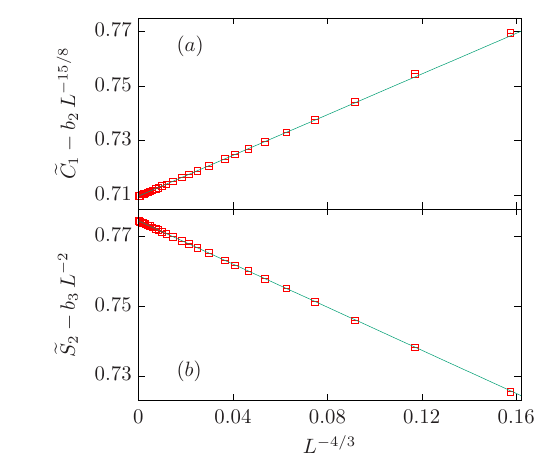}
\caption{\label{fig:Q2} 
FSS corrections for the O$(\sqrt{2})$ loop model on the dense branch, 
which has the same universality class as the critical Ising model. We show the 
quantities $\widetilde{C}_1 - b_2 L^{-15/8}$ with $b_2=-0.348$ in panel (a),
and  $\widetilde{S}_2 - b_3 L^{-2}$ with $b_3=0.496$ in panel (b). In both
panels, data is represented vs $L^{-4/3}$. 
It can be clearly seen that the coefficient of the leading correction, 
with exponent $-1/g$, has opposite signs for $C_1$ and $S_2$, suggesting 
that the small clusters contribute more than the largest one.
This scenario holds true for the whole dense branch till the O(2) case, 
where both contributions cancel each other.}
\end{figure}
%%%%%%%%%%%%%%%%%%%%%%%%%%%%%%%%%%%%%%%%%%%%%%%%%%%%%%

%%%%%%%%%%%%%%%%%%%%%%%%%%%%%%%%%%%%%%%%%%%%%%%%%%%%%%
% 
% TABLE 3 for Q=2 and Q=3
%
\begin{table*}[tb] 
\caption{\label{tab:Q=2-3}
FSS corrections for the O($\sqrt{2})$ and O($\sqrt{3})$ loop models on the 
dense branch, which belong to the same universality classes as the critical 
Ising model ($Q=2$) and the three-state Potts models, respectively. 
The former is characterized by $g=3/4$, $d_f=15/8$, and 
$2d_f-2 = 7/4$, and the latter is given by 
$g=5/6$, $d_f=28/15$, and $2d_f-2 = 26/15$. 
We show the fits of the quantities $\widetilde{C}_1$ and $\widetilde{S}_2$ 
to the full ansatz \eqref{def_ansatz_OK}.
}
\begin{ruledtabular}
\begin{tabular}{cccllllllll}
\multicolumn{11}{c}{Fits for the dense branch $x_-$} \\
\multicolumn{1}{c}{$n$} 
&\multicolumn{1}{c}{$\widetilde{O}$} 
&\multicolumn{1}{c}{$L_{\rm min}$} 
&\multicolumn{1}{c}{$a$} 
&\multicolumn{1}{c}{$b_1$}
&\multicolumn{1}{c}{$\omega$}
&\multicolumn{1}{c}{$b_2$} 
&\multicolumn{1}{c}{$y_2$}  
&\multicolumn{1}{c}{$b_3$}  
&\multicolumn{1}{c}{$y_3$}  
&\multicolumn{1}{c}{$\chi^2/{\rm DF}$} \\
\hline\\[-3mm]
$\sqrt{2}$ &$\widetilde{C}_1$ 
&40 &0.709702(6) &\kk0.258(8) &1.269(8)&\kk0     &     &\kk0     &  &0.56/7\\
& &\kz9 &0.709702(5) &\kk0.346(8) &1.317(5)&-0.30(1) &15/8 &\kk0     &  &10.3/18\\
& &10 &0.709701(5) &\kk0.316(7) &1.304(5)&\kk0     &     &-0.31(2) &2 &6.9/17\\
& &\kz9 &0.709703(4) &\kk0.383(4) &4/3     &-0.59(7) &15/8 &\kk0.27(8)&2 &8.3/18\\
& &10 &0.709711(3) &\kk0.3724(7)&4/3     &-0.348(3)&15/8 &\kk0     &  &9.5/18\\ \\[-2mm]
&$\widetilde{S}_2$ 
&28 &0.774568(9) & -0.179(6) &1.22(1) &\kk0      &     &\kk0       &  &6.3/10\\
& &12 &0.774550(9) & -0.47(5)  &1.39(2) &\kk0.55(6)&7/4  &\kk0       &  &7.0/15\\ 
& &\kz6 &0.774550(6) & -0.309(4) &1.332(4)&\kk0      &     &\kk0.494(7)&2 &15.2/21\\
& &\kz5 &0.774552(5) & -0.313(3) & 4/3    &\kk0.02(1)&7/4  &\kk0.47(1) &2 &16.7/22\\
& &\kz6 &0.774548(3) & -0.3098(3)& 4/3    &\kk0      &     &\kk0.496(1)&2 &15.3/22\\
%
% Q=3
%
\hline\\[-3mm]
$\sqrt{3}$ &$\widetilde{C}_1$ 
&24 &0.758498(9) &\kk0.192(3) &1.141(6)&\kk0     &     &\kk0     &  &12.6/11\\
& &10 &0.758516(9) &\kk0.258(9) &1.201(9)&-0.23(3) &28/15&\kk0     &  &14.8/17\\
& &10 &0.758512(9) &\kk0.237(7) &1.186(8)&\kk0     &     &-0.23(2) &2 &15.4/17\\
& &10 &0.758515(8) &\kk0.257(4) &6/5     &-0.2(1) &28/15&\kk0.0(1)&2 &14.9/17\\ 
& &10 &0.758515(5) &\kk0.2569(7)&6/5     &-0.226(3)&28/15&\kk0     &  &14.9/18\\ \\[-2mm]
&$\widetilde{S}_2$ 
&24   &0.82358(1) & -0.110(3) &1.084(9) &\kk0      &     &\kk0      &  &12.9/11\\
& &\kz7 &0.823531(9)& -0.265(9) &1.257(8) &\kk0.35(1)&26/15&\kk0      &  &16.5/20\\
& &\kz9 &0.82354(1) & -0.177(6) &1.191(9) &\kk0      &     &\kk0.32(2)&2 &12.9/18\\
& &\kz9 &0.82354(8) & -0.187(4) &6/5      &\kk0.04(4)&26/15&\kk0.29(5)&2 &13.1/18\\
& &\kz9 &0.823532(5)& -0.1831(5)&6/5      &\kk0      &     &\kk0.338(3)&2&13.9/19\\
\end{tabular}
\end{ruledtabular}
\end{table*}
%%%%%%%%%%%%%%%%%%%%%%%%%%%%%%%%%%%%%%%%%%%%%%%%%%%%%%

In this case, we have followed the procedure explained in Sec.~\ref{sec:Q=1};
the results are displayed in Table~\ref{tab:Q=2-3}. 

The first step consists in a simple power-law fit to the $\widetilde{C}_1$ data 
with $b_2=b_3=0$; the result for the exponent $\omega=1.269(8)$ is far away 
from the expected value $1/g=4/3$ (see Table~\ref{tab:Q=2-3}).
Better results for $\omega$ are obtained if we fix $y_2=15/8$ or $y_3=2$. The
fit with the three powers fixed to their expected values ($\omega=4/3$, 
$y_2=15/8$, and $y_3=2$) is rather good already for $L_\text{min}=9$. 
The value for the amplitude $b_3 = 0.27(8)$ is not far from zero (approximately
three times the error bar), and it has the same sign as $b_1$ but the 
opposite sign as $b_2$. Since $b_2$ and $b_3$ can influence each other 
and potentially cancel out in the fit, we retain the larger correction term 
$b_2$ and set $b_3 = 0$ in the final fit. Notice that we obtain two 
similar amplitudes with opposite signs: $b_1=0.3724(7)$ and $b_2=-0.348(3)$. 

The analysis of the $\widetilde{S}_2$ data follows the same procedure. 
In this case, the expected value for $\omega$ is very well reproduced if we
fix $b_2=0$ and $y_3=2$; namely $\omega=1.332(4)$. In the three-parameter fit 
with all powers fixed to their expected values, it is clear that $b_2=0$ 
within error. Therefore, our preferred fit corresponds to the last row of 
the second block in Table~\ref{tab:Q=2-3}. Again, we find two similar 
amplitudes with opposite signs: $b_1=-0.3098(3)$ and $b_3=0.496(1)$. 

In summary, we have found that both data sets can be well described with two
powers with exponents $1/g=4/3$ and $y_2=d_f$ ($y_3=2$) for the
$\widetilde{C}_1$ ($\widetilde{S}_2$) data set. In both cases, we
find two similar amplitudes with opposite signs (see Fig.~\ref{fig:Q2}).

Finally, it is worth noting that the correction exponent $1/g=4/3$ appears
only when considering geometric objects in the Ising model. In fact, for the
Ising model in the spin representation \cite{Salas_00}, one only finds 
integer exponents: i.e., $\omega=2$. This is exactly the expected behavior
\cite{Caselle_01,Caselle_02,Salas_01,Salas_02}. 

%%%%%%%%%%%%%%%%%%%%%%%%%%%%%%%%%%%%%%%%%%%%%%%%%%%%%%
%
% Q=3
%
%%%%%%%%%%%%%%%%%%%%%%%%%%%%%%%%%%%%%%%%%%%%%%%%%%%%%%
\subsubsection{\texorpdfstring{$n=\sqrt{3}$}{n=sqrt{3}}} \label{sec:Q=3}

%%%%%%%%%%%%%%%%%%%%%%%%%%%%%%%%%%%%%%%%%%%%%%%%%%%%%%
%
% FIGURE 5
%
\begin{figure}[htb]
\includegraphics[scale=0.9]{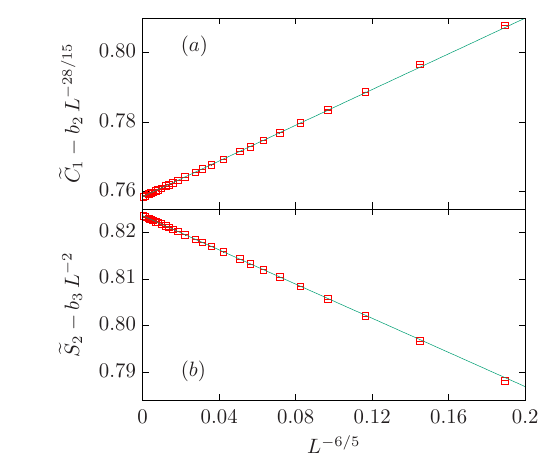}
\caption{\label{fig:Q3} 
FSS corrections for the O$(\sqrt{3})$ loop model on $x_-$, which
belongs to the same universality class as the critical three-state Potts model.
We show the quantities $\widetilde{C}_1 - b_2 L^{-28/15}$ with $b_2=-0.226$ 
in panel (a) and $\widetilde{S}_2 - b_3 L^{-2}$ with $b_3=0.338$ in panel (b).
In both panels, data is depicted vs $L^{-6/5}$. 
}
\end{figure}
%%%%%%%%%%%%%%%%%%%%%%%%%%%%%%%%%%%%%%%%%%%%%%%%%%%%%%

%%%%%%%%%%%%%%%%%%%%%%%%%%%%%%%%%%%%%%%%%%%%%%%%%%%%%%
% 
% TABLE 4 for Q=4 
%
\begin{table*}[tbh] 
\caption{\label{tab:Q=4}
FSS corrections for the O$(2)$ loop model and the O$(\sqrt{3})$ loop model 
on the dilute branch $x_+$, which belong to the same universality classes as 
the critical four-state Potts and the tricritical three-state Potts models, 
respectively. The former model is characterized by $g=1$, $d_f=15/8$, and 
$2d_f-2 = 7/4$, and the latter one, by $g=7/6$, $d_f=40/21$, 
$2d_f-2 = 38/21$, and $4/g-2=10/7$. We show the fits of the quantities 
$\widetilde{C}_1$ and $\widetilde{S}_2$ to the full ansatz 
\eqref{def_ansatz_OK}.
}
\begin{ruledtabular}
\begin{tabular}{ccllllllll}
\multicolumn{1}{c}{$\widetilde{O}$} 
&\multicolumn{1}{c}{$L_{\rm min}$} 
&\multicolumn{1}{c}{$a$} 
&\multicolumn{1}{c}{$b_1$}
&\multicolumn{1}{c}{$\omega$}
&\multicolumn{1}{c}{$b_2$} 
&\multicolumn{1}{c}{$y_2$}  
&\multicolumn{1}{c}{$b_3$}  
&\multicolumn{1}{c}{$y_3$}  
&\multicolumn{1}{c}{$\chi^2/{\rm DF}$} \\
\hline\\[-2.5mm]
\multicolumn{10}{c}{Fits for the O$(2)$ loop model} \\\\[-3mm]
$\widetilde{C}_1$
&24 &0.84673(3) &0.120(4) &0.97(1)&\kk0     &     &\kk0     &  &10.4/10\\
&\kz7 &0.84673(2) &0.127(4) &0.98(1)&-0.06(1) & 15/8&\kk0     &  &18.4/19\\
&\kz7 &0.84673(2) &0.124(3) &0.979(9)&\kk0    &     &-0.06(1) & 2&18.5/19\\
&\kz8 &0.84674(2) &0.137(2) &1       &-0.3(1) & 15/8&\kk0.3(1)& 2&16.2/18\\
&10 &0.84676(1) &0.1345(6)&1       &-0.082(5)& 15/8&\kk0    &  &15.5/17\\\\[-3mm]
$\widetilde{S}_2$
&9    &0.891851(4)	& 0.079(9)&2.01(5)&\kk0	     &     &\kk0     &  &16.1/18\\
&\kz8 &0.89183(1)  & 0.003(1)& 1     &-0.09(3)   & 7/4 &\kk0.20(5)& 2&13.5/18\\
&\kz9 &0.891850(4) & 0       &       &\kk0.000(8)& 7/4 &\kk0.08(1) & 2&16.1/18\\ 
&\kz9 &0.891850(3) & 0       &       &\kk0       &     &\kk0.0775(9)& 2&16.1/19\\
\hline\\[-2.5mm]
&\multicolumn{8}{c}{Fits for the O$(\sqrt{3})$ loop model on the dilute
branch $x_+$} \\\\[-3mm]
$\widetilde{C}_1$
&\kz8 &0.92423(3) &0.0610(6)&0.838(5)&\kk0      &      &\kk0     &  &14.6/19\\
&\kz6 &0.92431(4) &0.069(3) &0.88(1) &-0.029(6) & 40/21&\kk0     &  &11.6/20\\
&\kz6 &0.92430(4) &0.068(2) &0.87(1) &\kk0      &      &-0.031(8)& 2&11.6/20\\
&\kz6 &0.92428(2) &0.064(1) & 6/7    &\kk0.1(1) & 40/21&-0.2(1)  & 2&14.7/20\\
& 20 &0.92428(1) &0.0639(4)& 6/7    &\kk0      &      &\kk0     &  &5.52/12\\\\[-3mm]
$\widetilde{S}_2$
&\kz8 &0.95050(3) & 0.0542(7) &0.827(7)&\kk0     &     &\kk0      &  &13.2/19\\
&\kz4 &0.95059(3) & 0.064(2)  &0.88(1) &-0.031(4)&38/21&\kk0      &  &10.6/22\\
&\kz4 &0.95063(4) & 0.080(5)  &0.93(2) &-0.043(7)&10/7 &\kk0      &  &11.1/22\\
&\kz4 &0.95058(3) & 0.061(2)  &0.87(1) &\kk0     &     &-0.031(4) & 2&10.6/22\\
&\kz4 &0.95056(2) & 0.0590(8) & 6/7    &\kk0.02(2)&38/21&-0.05(3) & 2&10.7/22\\
&\kz4 &0.95057(2) & 0.059(1)  & 6/7    &\kk0.006(7)&10/7&-0.036(9)& 2&10.6/22\\
&\kz4 &0.95055(1) & 0.0597(2) & 6/7    &\kk0     &     &-0.028(1) & 2&11.4/23\\
\end{tabular}
\end{ruledtabular}
\end{table*}
%%%%%%%%%%%%%%%%%%%%%%%%%%%%%%%%%%%%%%%%%%%%%%%%%%%%%%

The O$(\sqrt{3})$ loop model on the dense branch belongs to the same 
universality class as the
critical three-state Potts model. Both models are characterized by $g=5/6$, 
$d_f=28/15 \approx 1.866667$, and $2d_f-2=26/15\approx 1.733333$.   

We have followed the procedure outlined in Sec.~\ref{sec:Q=1};
the results are displayed on the last two blocks of Table~\ref{tab:Q=2-3}. 

Again, if we fit the $\widetilde{C}_1$ data to a simple power-law with 
$b_2=b_3=0$, we obtain a result for the exponent $\omega=1.141(6)$ which 
is far away from the expected value $1/g=6/5$ (see Table~\ref{tab:Q=2-3}).
Better results for $\omega$ are obtained if we fix $y_2=28/15$ or $y_3=2$. The
fit with the three powers fixed to their expected values ($\omega=6/5$, 
$y_2=28/15$, and $y_3=2$) is rather good already for $L_\text{min}=10$. 
The value for the amplitude $b_3=0.0(1)$ agrees with zero within error, so we 
set it to $b_3=0$ in the last fit. We obtain two similar amplitudes
with opposite signs: $b_1=0.2569(7)$ and $b_2=-0.226(3)$.

The analysis of the $\widetilde{S}_2$ data follows the same procedure. 
In this case, the expected value for $\omega$ is very well reproduced if we
fix $b_2=0$ and $y_3=2$. In the three-parameter fit with all powers fixed to
their expected values, it is clear that $b_2=0$ within error bars. Therefore, 
our preferred fit corresponds to $b_1=-0.1831(5)$ and $b_3=0.338(3)$ (see 
Table~\ref{tab:Q=2-3}). They are quite similar with opposite signs. 

In summary, we have found that both data sets can be well described with two
powers with exponents $1/g=6/5$ and $y_2=d_f$ ($y_3=2$) for the
$\widetilde{C}_1$ ($\widetilde{S}_2$) data set. In both cases, we
find two similar amplitudes with opposite signs (see Fig.~\ref{fig:Q3}).

%%%%%%%%%%%%%%%%%%%%%%%%%%%%%%%%%%%%%%%%%%%%%%%%%%%%%%
%
% Q=4
%
%%%%%%%%%%%%%%%%%%%%%%%%%%%%%%%%%%%%%%%%%%%%%%%%%%%%%%
\subsection{\texorpdfstring{%
	O$\bm{(2)}$ loop model}{O(2) loop model}} \label{sec:Q=4}

The O$(2)$ loop model belongs to the same universality class as the critical 
four-state Potts model. Both models are characterized by $g=1$, 
$d_f=15/8 = 1.875$, and $2d_f-2=7/4= 1.75$. This model corresponds to the
point where the dense and dilute branches meet $x_-=x_+$. 

We have followed the procedure outlined in Sec.~\ref{sec:Q=1};
the results are displayed in Table~\ref{tab:Q=4}. 

In this case, one might expect logarithmic correction due to the existence
of the marginally irrelevant dilution field (i.e., $y_{t2}=0$)
\cite{Nauenberg_80,Cardy_80,Salas_97}. As observed in Ref.~\cite{Xu_25},
simulating the equivalent O$(2)$ loop model makes these logarithmic 
corrections to disappear.

\begin{figure}[htb]
\includegraphics[scale=0.9]{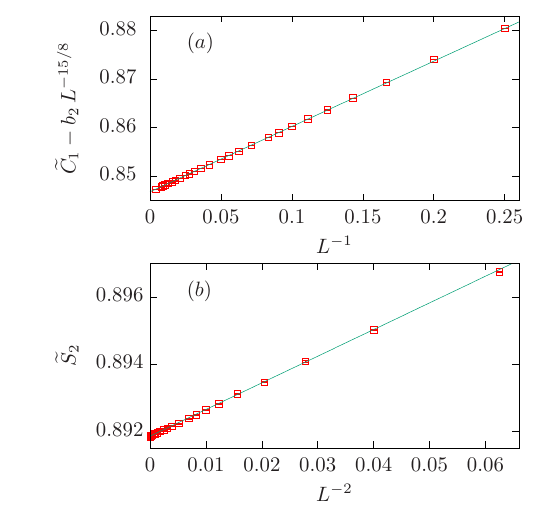}
\caption{\label{fig:Q4} 
FSS corrections for the O$(2)$ loop model, which belongs to the same
universality class as the critical four-state Potts model. We show the 
quantities $\widetilde{C}_1 - b_2 L^{-15/8}$ vs $L^{-1}$ 
with $b_2=-0.082$ in panel (a), and $\widetilde{S}_2$ vs $L^{-2}$ in panel (b). 
It is clearly seen that 
the theoretically predicted correction term, with exponent $-1$, 
is absent in $S_2$. This effect can be attributed to the exact cancellation 
of the contributions from large clusters and from the smaller ones. 
}
\end{figure}

The fit to the data $\widetilde{C}_1$ is straightforward. We find that the 
dominant exponent is $1/g=1$, while the subdominant exponent is 
$d_f=15/8$ [see Table~\ref{tab:Q=4} and Fig.~\ref{fig:Q4}(a)].

On the other hand, the fit to the data $\widetilde{S}_2$ presents some 
surprises. We first observe that the fit to a single power-law (i.e., setting
$b_2=b_3=0$) gives a power $\omega=2.01(5)$ with a small amplitude 
$b_1=0.079(9)$. Moreover, the fits with $y_2=d_f$, $b_3=0$ or with
$b_2=0$, $y_3=2$ give useless results: either the procedure does not converge,
or the error bars are extremely large. If we set $\omega=1/g=1$, 
$y_2=d_f$, and $y_3=2$, we find an extremely small value for the amplitude
$b_1$. Furthermore, if we set $b_1=0$, $y_2=d_f$, and $y_3=2$, the 
subsequent fit shows that $b_2=0$ within statistical errors. Therefore, for
this case we find that the dominant FSS term corresponds to $y_3=2$ with
a small amplitude $b_3$ [see Table~\ref{tab:Q=4} and Fig.~\ref{fig:Q4}(b)]. 

%%%%%%%%%%%%%%%%%%%%%%%%%%%%%%%%%%%%%%%%%%%%%%%%%%%%%%
%
% Q=1,2,3 TRICRITICAL
%
%%%%%%%%%%%%%%%%%%%%%%%%%%%%%%%%%%%%%%%%%%%%%%%%%%%%%%
\subsection{\texorpdfstring{%
	O$\bm{(n)}$ loop model on the dilute branch $\bm{x_+}$}
{O(n) loop model on the dilute branch}} \label{sec:Q=1-3T}

In this section we will discuss our results for O$(n)$ loop model on the
dilute branch $x_+$ with $n=\sqrt{3},\sqrt{2},1$, which belong to
the universality classes of the tricritical Potts model
with $Q=3,2,1$ states, respectively. 
The results are rather similar qualitatively, so we
will discuss the first case in more detail, and be brief in the other
two cases.  

%%%%%%%%%%%%%%%%%%%%%%%%%%%%%%%%%%%%%%%%%%%%%%%%%%%%%%
%
% FIGURE 6
%
%%%%%%%%%%%%%%%%%%%%%%%%%%%%%%%%%%%%%%%%%%%%%%%%%%%%%%

%%%%%%%%%%%%%%%%%%%%%%%%%%%%%%%%%%%%%%%%%%%%%%%%%%%%%%
%
% Q=3 TRICRITICAL
%
%%%%%%%%%%%%%%%%%%%%%%%%%%%%%%%%%%%%%%%%%%%%%%%%%%%%%%
\subsubsection{\texorpdfstring{$n=\sqrt{3}$}{n=sqrt{3}}} \label{sec:Q=3T}

%%%%%%%%%%%%%%%%%%%%%%%%%%%%%%%%%%%%%%%%%%%%%%%%%%%%%%
%
% FIGURE 7
%
\begin{figure}[htb]
\includegraphics[scale=0.9]{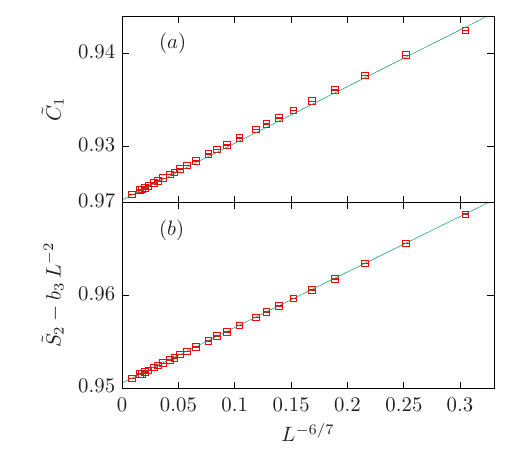}
\caption{\label{fig:qd3} 
FSS corrections for the O$(\sqrt{3})$ loop model on the dilute branch $x_+$,
which belongs to the same universality class as the tricritical three-state 
Potts model. We show the quantities $\widetilde{C}_1$ in panel (a) and
$\widetilde{S}_2- b_3 L^{-2}$ with $b_3=-0.028$ in panel (b). In
both panels, data points are depicted vs $L^{-6/7}$. 
}
\end{figure}
%%%%%%%%%%%%%%%%%%%%%%%%%%%%%%%%%%%%%%%%%%%%%%%%%%%%%%

%%%%%%%%%%%%%%%%%%%%%%%%%%%%%%%%%%%%%%%%%%%%%%%%%%%%%%
%
% FIGURE 8
%
\begin{figure}[hbt]
\includegraphics[scale=0.9]{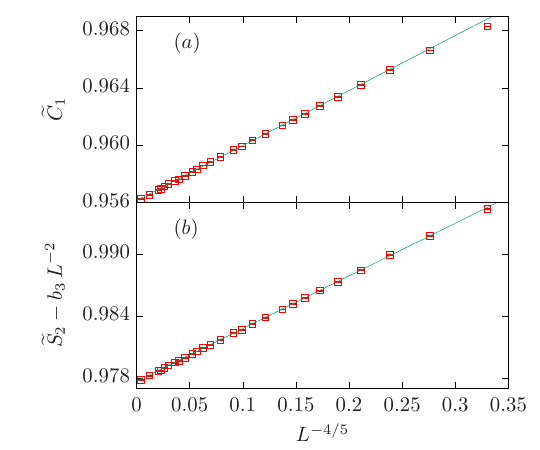}
\caption{\label{fig:qd2} 
FSS corrections for the O$(\sqrt{2})$ loop model on the dilute branch, which
belongs to the same universality class as the tricritical Ising model. 
We show the quantities $\widetilde{C}_1$ in panel (a) and
$\widetilde{S}_2- b_3 L^{-2}$ with $b_3=-0.039$ in panel (b). In both panels, 
we depict the data points vs $L^{-4/5}$. 
}
\end{figure}
%%%%%%%%%%%%%%%%%%%%%%%%%%%%%%%%%%%%%%%%%%%%%%%%%%%%%%

The O$(\sqrt{3})$ loop model on the dilute branch belongs to the same 
universality class as the tricritical three-state Potts model. Both
models are characterized by $g=7/6$, 
$d_f=40/21 \approx 1.904762$, $2d_f-2=38/21\approx 1.809524$, and 
$4/g-2=10/7 \approx 1.428571$.   

Let us start with the fit of the $\widetilde{C}_1$ data. If we fit the data to
a single power-law ansatz [e.g., \eqref{def_ansatz_OK}
with $b_2=b_3=0$], we obtain an estimate $\omega=0.838(5)$ which is far (i.e.
four standard deviations) from the expected result $1/g=6/7\approx 0.857143$.
Better estimates are obtained if we set $y_2=d_f=40/21$ and $b_3=0$, or
$b_2=0$ and $y_3=2$. If we set $\omega=1/g=6/7$, $y_2=d_f$, and $y_3=2$,
we obtain a good estimate for $L_\text{min}=6$. It is worth noting that the
estimates for both $b_2$ and $b_3$ are compatible with zero within two 
standard deviations. Finally, we set $\omega=1/g$, and $b_2=b_3=0$. The 
fit is reasonable for $L_\text{min}=20$ with a small (but nonzero) amplitude
$b_1$.  

%%%%%%%%%%%%%%%%%%%%%%%%%%%%%%%%%%%%%%%%%%%%%%%%%%%%%%
% 
% TABLE 5 for Q=2 and Q=1 TRICRITICAL
%
\begin{table*}[tb] 
\caption{\label{tab:Q=2-1T}
FSS corrections for the O$(\sqrt{2})$ and O(1) loop models 
on the dilute branch, which belong to the same universality class as 
the tricritical Ising model ($Q=2$) and the one-state
tricritical Potts model, respectively. The former is characterized by $g=5/4$,
$d_f=77/40$, $2d_f-2=37/20$, and $4/g-2=6/5$. The latter is
given by $g=4/3$, $d_f=187/96$, $2d_f-2=91/48$, and $4/g-2=1$.
We show the fits of the quantities $\widetilde{C}_1$ and $\widetilde{S}_2$ 
to the full ansatz \eqref{def_ansatz_OK}.
}
\begin{ruledtabular}
\begin{tabular}{cccllllllll}
\multicolumn{11}{c}{Fits for the dilute branch $x_+$} \\
\multicolumn{1}{c}{$n$} 
&\multicolumn{1}{c}{$\widetilde{O}$} 
&\multicolumn{1}{c}{$L_{\rm min}$} 
&\multicolumn{1}{c}{$a$} 
&\multicolumn{1}{c}{$b_1$}
&\multicolumn{1}{c}{$\omega$}
&\multicolumn{1}{c}{$b_2$} 
&\multicolumn{1}{c}{$y_2$}  
&\multicolumn{1}{c}{$b_3$}  
&\multicolumn{1}{c}{$y_3$}  
&\multicolumn{1}{c}{$\chi^2/{\rm DF}$} \\
\hline\\[-3mm]
$\sqrt{2}$ &$\widetilde{C}_1$
&\kk5 &0.95603(2) &0.0372(3) &0.779(4)&\kk0       &     &\kk0    &  &15.4/23\\
& &\kk5 &0.95604(3) &0.037(1)  &0.78(1) &\kk0.00(4) &77/40&\kk0    &  &15.4/22\\
& &\kk5 &0.95604(3) &0.037(1)  &0.78(1) &\kk0       &     &\kk0.00(4)&2&15.4/22\\
& &\kk5 &0.95606(2) &0.0401(6) & 4/5    &-0.13(8)   &77/40&\kk0.13(9) &2&15.4/22\\
& &20 &0.95606(2) &0.0396(4) & 4/5    &\kk0       &     &\kk0       & &9.60/13\\\\[-2mm]
&$\widetilde{S}_2$
& 20 &0.97761(5)&0.050(3) &0.79(2) &\kk0      &     &\kk0      &  &10.4/12\\
& &\kz5  &0.97760(4)&0.051(2) &0.79(1) &-0.027(5) &37/20&\kk0      &  &16.8/22\\
& &\kz8  &0.97763(6)&0.08(2)  &0.86(5) &-0.04(2)  & 6/5 &\kk0      &  &13.7/19\\
& &\kz5  &0.97759(4)&0.049(2) &0.79(1) &\kk0      &     &-0.029(6) & 2&16.9/22\\
& &\kz6  &0.97759(3)&0.053(1) & 4/5    &-0.14(7)  &37/20&\kk0.14(9)& 2&14.3/21\\
& &\kz6  &0.97758(3)&0.055(2) & 4/5    &-0.013(8) & 6/5 &-0.01(1)  & 2&15.0/21\\
& &\kz7  &0.97762(2)&0.0516(4)& 4/5    &\kk0      &     &-0.039(4) & 2&15.6/21\\
%
% Q=1
%
\hline\\[-3mm]
1 &$\widetilde{C}_1$
&10 &0.98067(3)&0.0174(6)&0.72(2)&\kk0       &      &\kk0       &  &16.4/18\\
& &\kz5 &0.98069(4)&0.018(1) &0.73(2)&\kk0.000(4)&187/96&\kk0       &  &21.2/22\\
& &\kz5 &0.98069(3)&0.018(1) &0.73(2)&\kk0       &      &\kk0.000(4)&2 &21.2/22\\
& &\kz8 &0.98068(3)&0.0200(9)& 3/4   &-0.5(3)    &187/96&\kk0.5(3)  &2 &18.8/19\\
& &14 &0.98071(1)&0.0187(2)& 3/4   &\kk0       &      &\kk0       &  &15.7/16\\\\[-3mm]
&$\widetilde{C}_1$
&14 &0.99850(6)&0.027(1) &0.70(2) &\kk0      &     &\kk0      &  &13.6/15\\
& &\kz6 &0.99854(6)&0.030(2) &0.73(3) &-0.021(7) &91/48&\kk0      &  &21.1/21\\
& &\kz4 &0.99857(5)&0.032(1) &0.75(2) &\kk0      &     &-0.033(4) & 2&22.8/22\\
& &\kz4 &0.99858(3)&0.0307(6)&3/4     &\kk0.04(5)&91/48&-0.07(6)  & 2&22.3/23\\
& &\kz4 &0.99858(4)&0.030(2) & 3/4    &\kk0.001(4)&1   &-0.034(5) & 2&22.7/23\\
& &\kz4 &0.99856(2)&0.0312(2)& 3/4    &\kk0      &     &-0.032(1) & 2&22.8/24\\
\end{tabular}
\end{ruledtabular}
\end{table*}
%%%%%%%%%%%%%%%%%%%%%%%%%%%%%%%%%%%%%%%%%%%%%%%%%%%%%%

Dealing with the $\widetilde{S}_2$ is now a bit more complicated than for
its dense-branch counterpart, as there is one extra candidate that may play a 
role; namely the exponent $4/g-2=10/7$. As before, we start with a single 
power-law fit [$b_2=b_3=0$ in \eqref{def_ansatz_OK}]. The result is 
$\omega=0.827(7)$, which is again many standard deviations from the expected 
result. Better results in average can be obtained by fixing one 
correction-to-scaling exponent: either $y_2=2d_f-2$, $y_2=4/g-2$, or $y_3=2$. 
It is striking that these fits are excellent already for $L_\text{min}=4$. 
If we now fix $\omega=1/g=6/7$, $y_3=2$, and either $y_2=2d_f-2$ or 
$y_2=4/g-2$, we obtain good fits with the amplitude $b_2=0$ within errors.
The amplitude $b_3$ seems to be small, but nonzero. We can check this 
observation by fixing $\omega=1/g$, $b_2=0$, and $y_3=2$. The result of
such a fit confirms the previous observation. It is approximately
half of $b_1$ with the opposite sign. 

In summary, we have found that the $\widetilde{C}_1$ data set can be
well described with a single power with exponent $1/g=6/7$. The amplitude
$b_1$ is rather small compared to $a$. On the other hand, the 
$\widetilde{S}_2$ data set can be described with two exponents 
$1/g=6/7$ and $y_3=2$. The amplitudes $b_1$ and $b_3$ are rather similar
in absolute value, but have opposite signs. We have already found this 
situation in the study of the dense branch. These two scenarios are depicted 
in Fig.~\ref{fig:qd3}. 

%%%%%%%%%%%%%%%%%%%%%%%%%%%%%%%%%%%%%%%%%%%%%%%%%%%%%%
%
% Q=2 TRICRITICAL
%
%%%%%%%%%%%%%%%%%%%%%%%%%%%%%%%%%%%%%%%%%%%%%%%%%%%%%%
\subsubsection{\texorpdfstring{$n=\sqrt{2}$}{n=sqrt{2}}} \label{sec:Q=2T}

The O$(\sqrt{2})$ loop model on the dilute branch belongs to the same
universality class as the tricritical Ising model ($Q=2$). Both models are 
characterized by $g=5/4$, 
$d_f=77/40 = 1.925$, $2d_f-2=37/20= 1.85$, and $4/g-2=6/5=1.2$.   

Let us start with the fit of the $\widetilde{C}_1$ data. If we fit the data to
a single power-law ansatz [i.e., $b_2=b_3=0$ in the ansatz 
\eqref{def_ansatz_OK}], we obtain an estimate $\omega=0.779(4)$ which is
five standard deviations away from the expected result $1/g=4/5= 0.8$
(see Table~\ref{tab:Q=2-1T}).
If we set $y_2=d_f,b_3=0$ or $y_3=2,b_2=0$, we obtain a similar result 
for $\omega$, as
the estimates for $b_2$ or $b_3$ are compatible with zero within error bars.
Therefore, we refit the data assuming that $b_2=b_3=0$, and obtain a good
fit for a larger value of $L_\text{min}=20$. The plot of $\widetilde{C}_1$ vs
$L^{-4/5}$ is depicted in Fig.~\ref{fig:qd2}(a). 

The fit to the $\widetilde{S}_2$ data is similar. The initial fit to a simple
power law (see Table~\ref{tab:Q=2-1T}) shows an stable fit for 
$L_\text{min}=20$ which yields an estimate for $\omega=0.79(2)$ that is
compatible within error bars with the expected result $1/g=4/5= 0.8$.  
The fits with either $b_3=0,y_2=2d_f-2,4/g-2$ or $b_2=0,y_3=2$ 
fixed give estimates for
$\omega$ that are compatible within error bars with the expected result. It
is worth noticing that the fit with $y_2=4/g-2=6/5$ is rather unstable 
numerically. The amplitudes $b_2$ and $b_3$ are rather small. We chose $b_2=0$
in our final estimate as in previous cases. The data is
displayed in Fig.~\ref{fig:qd2}(b). 

%%%%%%%%%%%%%%%%%%%%%%%%%%%%%%%%%%%%%%%%%%%%%%%%%%%%%%
%
% Q=1 TRICRITICAL
%
%%%%%%%%%%%%%%%%%%%%%%%%%%%%%%%%%%%%%%%%%%%%%%%%%%%%%%
\subsubsection{\texorpdfstring{$n=1$}{n=1}} \label{sec:Q=1T}

The O$(1)$ loop model on the dilute branch $x_+$ is just the triangular-lattice
Ising model, and the clusters studied hereby are the Ising domains. The 
critical behavior of these Ising domains is expected to belong to the same 
universality class as the tricritical one-state Potts model. Both models are
characterized by $g=4/3$, 
$d_f=187/96 \approx 1.9479167$, $2d_f-2=91/48\approx 1.895833$,
and $4/g-2=1$. 

%%%%%%%%%%%%%%%%%%%%%%%%%%%%%%%%%%%%%%%%%%%%%%%%%%%%%%
%
% FIGURE 9
%
\begin{figure}[tbh]
\includegraphics[scale=0.9]{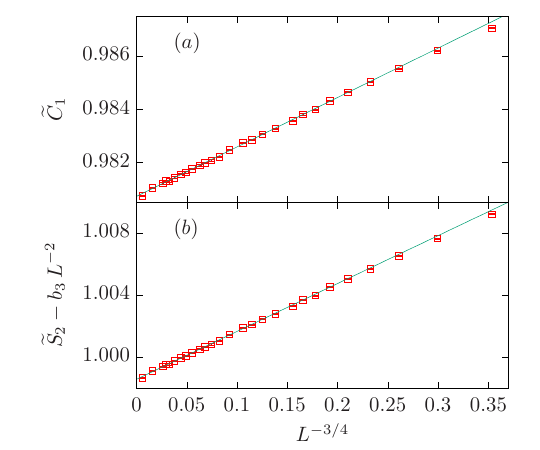}
\caption{\label{fig:qd1} 
FSS corrections for the O$(1)$ loop model on the dilute branch, which belongs 
to the same universality class as the tricritical one-state Potts model. 
We show the quantities $\widetilde{C}_1$ in panel (a) and
$\widetilde{S}_2-b_3 L^{-2}$ with $b_3=-0.032$ in panel (b). In both
panels, data points are depicted vs $L^{-3/4}$. 
}
\end{figure}
%%%%%%%%%%%%%%%%%%%%%%%%%%%%%%%%%%%%%%%%%%%%%%%%%%%%%%

Let us start with the fit of the $\widetilde{C}_1$ data. If we fit the data to
a single power-law ansatz [i.e., $b_2=b_3=0$ in the ansatz 
\eqref{def_ansatz_OK}], we obtain an estimate $\omega=0.72(2)$ which is
$1.5$ standard deviations away from the expected result $1/g=3/4= 0.75$
(see Table~\ref{tab:Q=2-1T}).
If we set $y_2=d_f,b_3=0$ or $y_3=2,b_2=0$, we obtain a similar result 
for $\omega$, as
the estimates for $b_2$ or $b_3$ are compatible with zero within error bars.
Therefore, we refit the data assuming that $b_2=b_3=0$, and obtain a good
fit for a larger value of $L_\text{min}=14$. The plot of $\widetilde{C}_1$ vs
$L^{-3/4}$ is depicted in Fig.~\ref{fig:qd1}(a). 

The fit to the $\widetilde{S}_2$ data is similar. The initial fit to a simple
power law (see Table~\ref{tab:Q=2-1T}) shows an stable fit for 
$L_\text{min}=14$, which yields an estimate for $\omega=0.70(2)$ that is
$2.5$ standard deviations from the expected result $1/g=3/4= 0.75$.  
The fits with either $b_3=0,y_2=2d_f-2$ or $b_2=0,y_3=2$ 
give estimates for
$\omega$ that are compatible within error bars with the expected result. It
is worth noticing that the fit with $y_2=4/g-2=1$ is very unstable 
numerically, and has not been included in Table~\ref{tab:Q=2-1T}. 
The amplitudes $b_2$ and $b_3$ are rather small. We chose $b_2=0$
in our final estimate as in previous cases. The data is
displayed in Fig.~\ref{fig:qd1}(b). 

Notice that the O$(1)$ loop model on the dilute branch has the same conformal
charge as the critical Ising model ($c=1/2$). 
However, since we study the critical behavior of the geometric Ising domains 
instead of the conventional thermodynamic quantities, 
the FSS corrections are very different, as we have seen in this section.

%%%%%%%%%%%%%%%%%%%%%%%%%%%%%%%%%%%%%%%%%%%%%%%%%%%%%%
%
% DISCUSSION 
%
%%%%%%%%%%%%%%%%%%%%%%%%%%%%%%%%%%%%%%%%%%%%%%%%%%%%%%
\section{Discussion} \label{sec:discussion}

We have studied the FSS corrections for geometric observables in the 
O$(n)$ loop model on both the dense and dilute branches. These models 
belong to the same universality classes as the critical and tricritical 
Potts models with $Q=n^2$. We have found that
the leading correction term has an exponent given by $1/g$ in agreement with 
previous studies \cite{Asikainen_03,Aharony_03}. We have also found that for
the quantity $C_1$, the second correction term has an exponent $d_f$.  
For the quantity $S_2$, the subleading correction term has an 
exponent $2$. No numerical evidence is obtained for the predicted 
subleading correction exponent $4/g-2$ for the dilute branch $x_+$ of the 
O$(n)$ loop model (i.e., the tricritical $Q=n^2$ Potts model), 
which might be due to some symmetries in the O$(n)$ loop model.

%%%%%%%%%%%%%%%%%%%%%%%%%%%%%%%%%%%%%%%%%%%%%%%%%%%%%%
%
% FIGURE 10 
%
\begin{figure}[htb]
\centering
\includegraphics[scale=0.95]{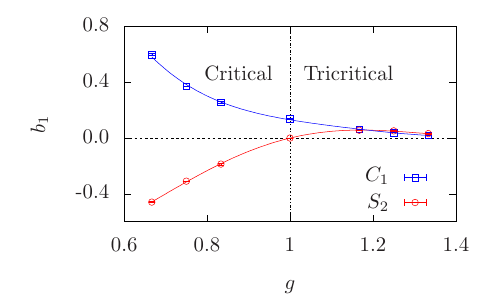}
\caption{\label{fig:b1}
Coefficient $b_1$ associated to the term $L^{-\omega}$ with $\omega=1/g$ 
in the ansatz \eqref{def_ansatz_OK}. 
We display the numerical results for $b_1$ obtained 
from the fits to $C_1$ (blue curve) and $S_2$ (red curve). 
Curves are only to guide the eye.}
\end{figure}
%%%%%%%%%%%%%%%%%%%%%%%%%%%%%%%%%%%%%%%%%%%%%%%%%%%%%%

We have also studied the amplitudes of the corresponding terms. In general,
we have found that the amplitudes of the leading and subleading terms are
similar and with opposite signs. This is the reason why extracting these
exponents has been very elusive in previous studies. 
The amplitude of the leading term $b_1$ is the one with smallest error bars,
so we can draw some firmer conclusions. 
In Fig.~\ref{fig:b1}, we have plotted the
amplitude $b_1$ as a function of the CG coupling $g$ for all the models we
have considered in this paper. For $C_1$ (blue curve in Fig.~\ref{fig:b1}), 
$b_1$ is positive and seems to be a decreasing function of $g$ in the interval 
$g\in [2/3,4/3]$, ranging from $b_1=0.602(2)$ for $g=2/3$ to 
$b_1= 0.0187(2)$ for $g=4/3$.
On the other hand, the amplitude $b_1$ for $S_2$ (red curve in 
Fig.~\ref{fig:b1}) is negative for $g\in [2/3,1)$, vanishes at $g=1$, 
and becomes positive and small for $g\in (1,4/3]$. In this latter regime,
it is quite similar to the amplitude $b_1$ for $C_1$. 
We recall here the fact (explained in Sec.~\ref{sec:fss}) that the amplitude
$b_1$ for $S_2$ has contributions from the amplitude $b_1$ for $C_1$ (large
clusters) and from the size distribution $n(s,p_c)$ (small clusters). This
picture explains the behavior shown in Fig.~\ref{fig:b1}: the red curve 
contains a (negative) contribution from small clusters and a (positive)
contribution from large clusters (which is given by the blue curve). 
As $g$ increases, these two contributions become to approximate in 
absolute value, and exactly at $Q=4$, they compensate so $b_1=0$ for $S_2$. 
The amplitude $b_2$ of the subleading term is similar to and with opposite
sign than $b_1$. Moreover, $b_2$ for $C_1$ is zero for the tricritical 
Potts models. 

Our study demonstrates the existence of the subleading 
magnetic scaling field in both the finite-cluster-size and 
finite-system-size corrections. It delivers a warning message that, in 
FSS analysis of the Potts model, the frequently ignored contributions 
from the subleading magnetic field might play a non-negligible
contributions. It would be interesting to extend our studies 
for the percolation and the Ising model in three and higher dimensions, where 
little knowledge has been known so far.

\begin{acknowledgments} 
This work has been supported by the National Natural Science Foundation of
China (under Grant No. 12275263), the Innovation Program for Quantum Science
and Technology (under grant No. 2021ZD0301900), the Natural Science 
Foundation of Fujian Province of China (under Grant No. 2023J02032).
\end{acknowledgments}

%\nocite{*}
%\bibliographystyle{apsrev4-1}

%\bibliography{ref}
 \input{fss_v4.bbl}
\end{document}